\documentclass[pdflatex,sn-mathphys-num]{sn-jnl}% Math and Physical Sciences Numbered Reference Style
\usepackage{amsmath,mathrsfs}
\usepackage{graphicx}%
\usepackage{multirow}%
\usepackage{amsmath,amssymb,amsfonts}%
\usepackage{amsthm}%
\usepackage{mathrsfs}%
\usepackage[title]{appendix}%
\usepackage{xcolor}%
\usepackage{textcomp}%
\usepackage{manyfoot}%
\usepackage{booktabs, threeparttable, tablefootnote}%
\usepackage{algorithm}%
\usepackage{algorithmicx}%
\usepackage{algpseudocode}%
\usepackage{listings}%
\usepackage{rotating}
\usepackage{graphicx}
\usepackage{adjustbox}
\theoremstyle{thmstyleone}%
\theoremstyle{thmstyletwo}%

\theoremstyle{thmstylethree}%

\newcommand{\resultNumAGNPlotted}{7101}
\newcommand{\resultNumAGNFitted}{6992}
\newcommand{\resultAgeUniverse}{\ensuremath{11.26\pm0.20\,\mathrm{Gyr}}}

\newcommand{\resultLogZFlatwZerowaCDM}{5460.6}
\newcommand{\resultLogZerrFlatwZerowaCDM}{0.2}

\newcommand{\resultSigmaFlatwCDM}{3.9}

\newcommand{\resultSigmaFlatLambdaCDM}{4.8}

\newcommand{\resultDeltaLogZFlatwZeroWaCDMFlatwCDM}{\ensuremath{7.5 \pm 0.2}}
\newcommand{\resultSigmaFlatwZeroWaCDMFlatwCDM}{3.9}

\newcommand{\resultDeltaLogZFlatwZeroWaCDMFlatLambdaCDM}{\ensuremath{11.5 \pm 0.2}}
\newcommand{\resultSigmaFlatwZeroWaCDMFlatLambdaCDM}{4.8}

\newcommand{\resultSigmaUVPivot}{\ensuremath{0.2}}
\newcommand{\resultTauUVRFPivot}{\ensuremath{800}}
\newcommand{\resultOmZero}{\ensuremath{0.57 \pm 0.02}}
\newcommand{\resultwZero}{\ensuremath{-0.46 \pm 0.40}}
\newcommand{\resultwa}{-16.44 \pm 5.70}
\newcommand{\resultLIntercept}{\ensuremath{(9.3 \pm 0.4)\times 10^{44}\,\mathrm{erg\,s^{-1}}}}
\newcommand{\resultAlphaAGNL}{\ensuremath{-2.52 \pm 0.06}}
\newcommand{\resultBetaAGNL}{\ensuremath{0.62 \pm 0.033}}
\newcommand{\resultScatterHD}{\ensuremath{0.81 \pm 0.023\,\mathrm{mag}}}
\newcommand{\resultScatterL}{\ensuremath{0.322 \pm 0.009\,\mathrm{dex}}}

\newcommand{\resultzonecutNumAGNFitted}{5648}

\newcommand{\resultzonecutSigmaFlatwCDM}{3.8}

\newcommand{\resultzonecutSigmaFlatLambdaCDM}{4.4}

\newcommand{\resultzonecutSigmaFlatwZeroWaCDMFlatwCDM}{3.8}

\newcommand{\resultzonecutSigmaFlatwZeroWaCDMFlatLambdaCDM}{4.4}

\begin{document}

\title[Evidence for evolving Dark Energy from a new cosmic probe]{Evidence for evolving Dark Energy from a new cosmic probe}


\author[1,2]{\fnm{Isaque} \sur{Dutra}}\email{isaque.dutra@yale.edu}
\equalcont{These authors contributed equally to this work.}

\author*[2]{\fnm{Colin J.} \sur{Burke}}\email{colin.j.burke@yale.edu}
\equalcont{These authors contributed equally to this work.}

\author*[1,2,3]{\fnm{Priyamvada} \sur{Natarajan}}\email{priyamvada.natarajan@yale.edu}

\author[4]{\fnm{Weixiang} \sur{Yu}}\email{wyu@UBishops.ca}

\affil[1]{\orgdiv{Department of Physics}, \orgname{Yale University}, \orgaddress{\street{217 Prospect Street}, \city{New Haven}, \state{CT} \postcode{06511}, \country{USA}}}

\affil[2]{\orgdiv{Department of Astronomy}, \orgname{Yale University}, \orgaddress{\street{219 Prospect Street}, \city{New Haven}, \state{CT} \postcode{06511}, \country{USA}}}

\affil[3]{\orgdiv{Black Hole Initiative}, \orgname{Harvard University}, \orgaddress{\street{20 Garden Street}, \city{Cambridge}, \state{MA} \postcode{02138}, \country{USA}}}

\affil[4]{\orgdiv{Department of Physics \& Astronomy}, \orgname{Bishop's University}, \orgaddress{\street{2600 College St}, \city{Sherbrooke}, \postcode{QC J1M 1Z7}, \country{Canada}}}
%TC:endignore


\abstract{The $\Lambda$CDM concordance cosmological model provides a remarkably successful description of the formation and evolution of structure in the Universe. However, a growing discrepancy between measurements of the expansion rate $H_0$ from the near and distant Universe now appears to be significant at the $\sim4{-}7\sigma$ level \citep{DiValentino2021,H0DNCollaboration2025}. This inconsistency, known as the ``Hubble tension'', has arisen either due to unrecognized systematics in these measurements or new physics beyond the standard model, such as an evolving dark energy equation of state \citep{MarcKAdamR2023,Freedman2023}. 
Modeling $\sim 20$-year, multi-band optical light curves for $\resultNumAGNFitted$ active galactic nuclei (AGN), we find a tight relation linking the variability amplitude and characteristic timescale to their intrinsic luminosity. This empirical law enables us to construct an AGN-based Hubble diagram to $z \sim 3.5$. Joint inference with supernova distances reveals evidence for an evolving dark energy equation of state at the $\resultzonecutSigmaFlatwZeroWaCDMFlatwCDM{-}\resultSigmaFlatwZeroWaCDMFlatwCDM\sigma$ over constant-$w$ models and $\resultzonecutSigmaFlatwZeroWaCDMFlatLambdaCDM{-}\resultSigmaFlatwZeroWaCDMFlatLambdaCDM\sigma$ over $\Lambda$CDM.
Our results establish AGN light curves as a powerful tool for cosmography that could offer a novel pathway to test deviations from the standard $\Lambda$CDM expansion history.}

%\keywords{keyword1, Keyword2, Keyword3, Keyword4}



\maketitle

Rapidly accreting supermassive black holes, known as active galactic nuclei (AGN), produce ultraviolet/optical (UV/optical) flux variations that are believed to trace temperature fluctuations in their accretion disks \citep{Shakura1973}. Although the physical mechanisms that drive these flux variations remain uncertain, empirical correlations between UV/optical variability and fundamental properties of AGNs, such as black hole mass and luminosity, are well known \citep{Hawkins2000,Wilhite2008,Kelly2009,MacLeod2010,Simm2016,Suberlak2021,Burke2021,Tarrant2025,BenatiGoncalves2025}. However, limitations due to the sampled duration of these light curves (flux vs. time); wavelength-dependent effects; contamination from variability outside the accretion disk; and various selection biases have rendered previous attempts to standardize these relations uncertain, hindering their potential use as distance indicators. We report on the existence of a tight correlation between the UV variability parameters of the rest-frame (RF) and the luminosity of the AGNs using multi-band modeling of $\sim20$ year-long light curves of AGNs in the redshift range $z\sim0.5-3.5$, consistent with previous work \citep{Suberlak2021}. Using this correlation, we construct a variable AGN distance -- luminosity relation, known as a \emph{Hubble diagram}. We demonstrate that AGN variability offers a powerful new tool to probe the cosmic expansion history and permits independent testing of dark energy models. Our method provides $\resultzonecutSigmaFlatwZeroWaCDMFlatwCDM{-}\resultSigmaFlatwZeroWaCDMFlatwCDM\sigma$ evidence for an evolving dark energy equation of state over constant-$w$ models and $\resultzonecutSigmaFlatwZeroWaCDMFlatLambdaCDM{-}\resultSigmaFlatwZeroWaCDMFlatLambdaCDM\sigma$ evidence over $\Lambda$CDM.

Various methods have been employed to constrain the properties of dark energy, including baryon acoustic oscillations (BAO; e.g., \citep{Dawson2016,Ross2017}); cosmic microwave background (CMB) measurements (e.g., \citep{PlanckCollaboration2020}); weak gravitational lensing (e.g., \citep{Abbott2022}); cluster strong lensing (e.g., \citep{Jullo2010,Gilmore2009}), and Type Ia supernovae (SN Ia; e.g., \citep{Riess2004,Wood-Vasey2007,Conley2011,Suzuki2012}). Recent findings from the Dark Energy Spectroscopic Instrument (DESI) using BAO, hint at possible redshift evolution in the dark energy equation of state at the $2.8{-}4.2\sigma$ significance level \citep{DESICollaboration2025}, highlighting the need for independent tests using complementary cosmological probes.  Given their high intrinsic luminosities of $L_{\rm{bol}} \sim 10^{44}{-}10^{48}$ erg~s$^{-1}$, AGNs act as cosmic lighthouses, enabling distance measurements to more than four times beyond those reachable by SN Ia, a robust standard candle. At these distances, differences between competing cosmological models become significantly more distinguishable. The existence of a correlation between AGN luminosity and rest-frame UV variability that enables their standardization could be explained by models in which the amplitude and timescale of thermal fluctuations in their accretion disks scale with both their accretion disk sizes and mass accretion rates.

\begin{figure}
\centering
\includegraphics[width=\textwidth]{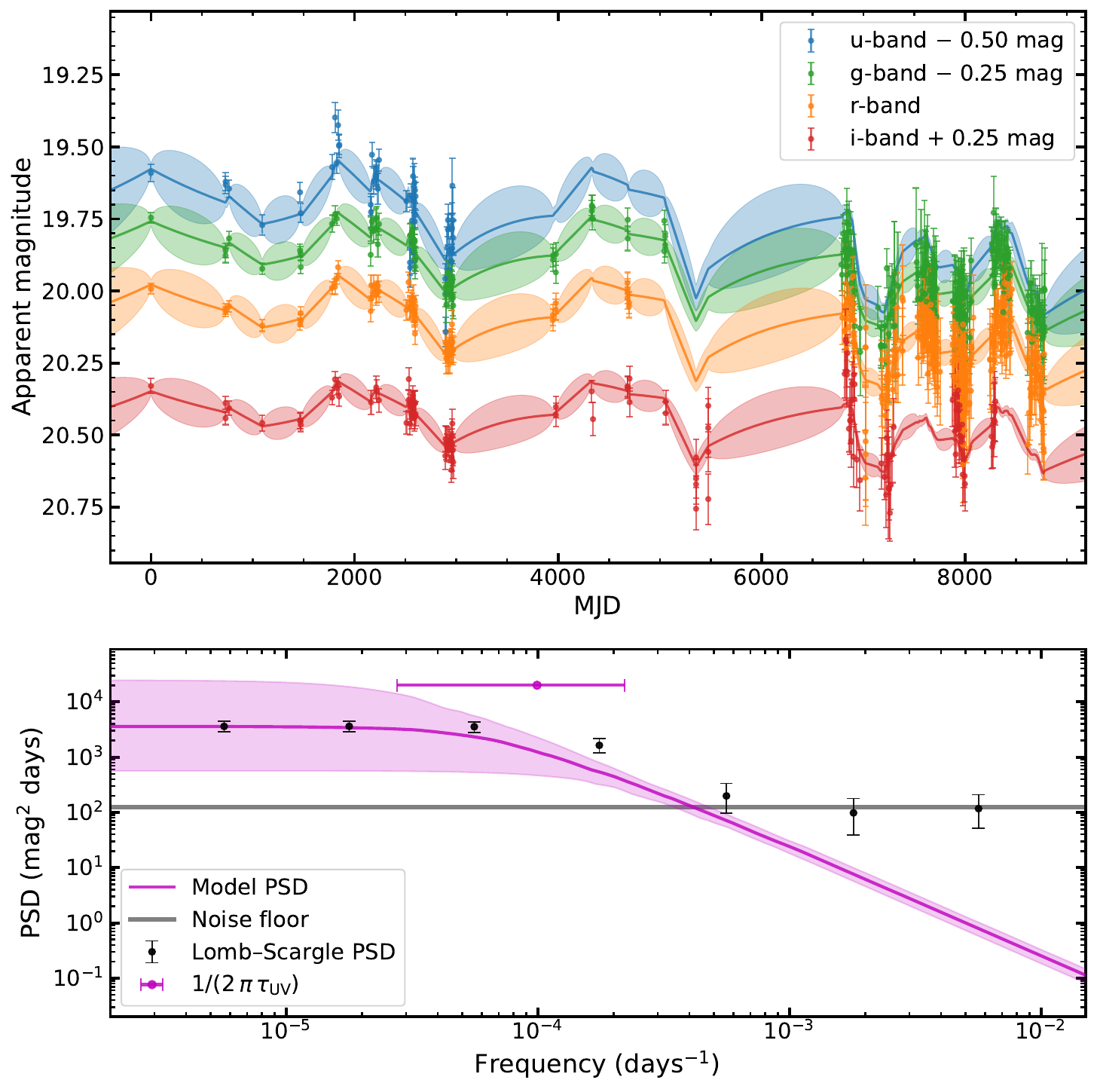}
\caption{\emph{Upper panel:} An illustrative example of an AGN light curve from our sample and the fitted multi-band model. The light curve in each band is offset by a scale factor for visual clarity. \emph{Lower panel:} Corresponding multi-band power spectral density (PSD) in angular frequencies from the model (magenta) and computed directly from the data using the Lomb-Scargle algorithm (black points; \citep{Lomb1976,Scargle1982}). The solid gray line is the noise floor level. The magenta error bar area is the 1$\sigma$ range of the angular frequency corresponding to the UV damping timescale. \label{fig:lc}} 
\end{figure}

\begin{figure}
\centering
\includegraphics[width=.8\textwidth]{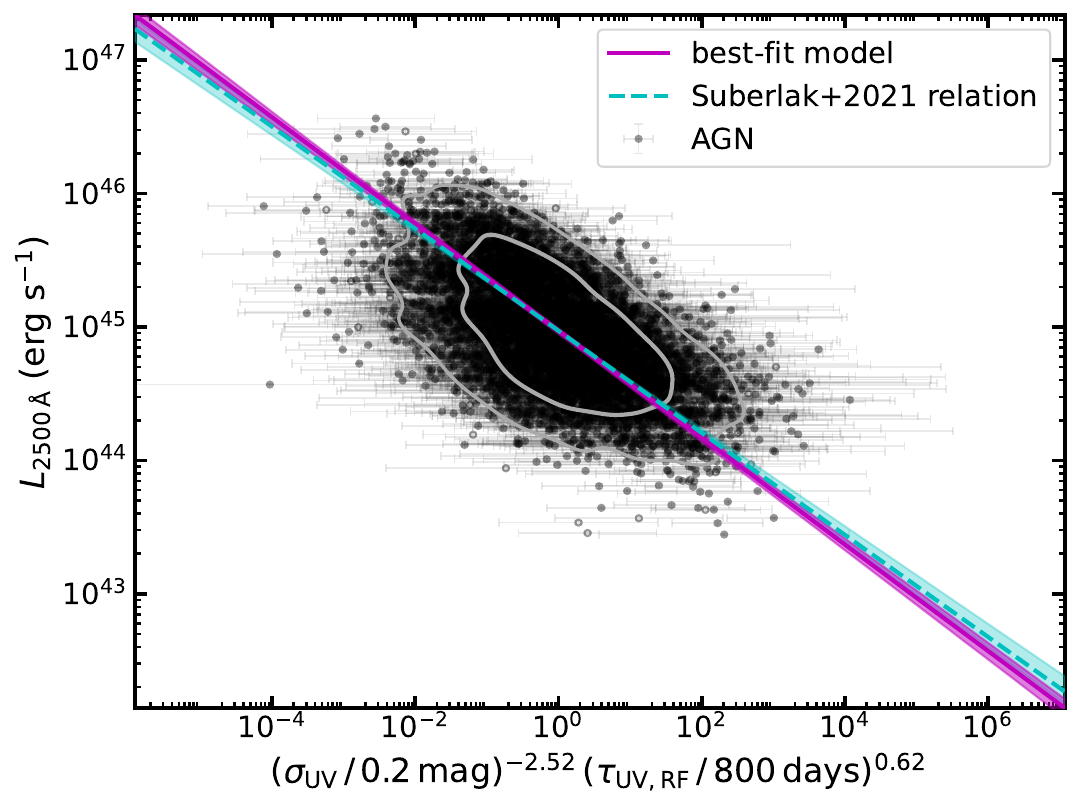}
\caption{Empirical relation between the bias-corrected rest-frame 2500 \AA\ monochromatic AGN luminosity, $L_{2500\,\text{\AA}}$, and a power-law combining the variability amplitude and timescale parameters. The best-fit model is shown in magenta along with the $1\sigma$ uncertainty band. The filled circles are AGN where rest-frame $2500$ \AA\ is within the wavelength coverage of the spectrum. The open circles denote sources where rest-frame $2500$ \AA\ is beyond the wavelength coverage of the spectrum, requiring an extrapolation of the AGN continuum. Open circles are not fitted, but are included in the figure for completeness. The slope of the best-fit relation with uncertainties from \citep{Suberlak2021} is shown as the cyan dashed line. The gray curves are 1 and 2$\sigma$ density contours. \label{fig:l2500}}
\end{figure}

\begin{figure}
\centering
\includegraphics[width=1\textwidth]{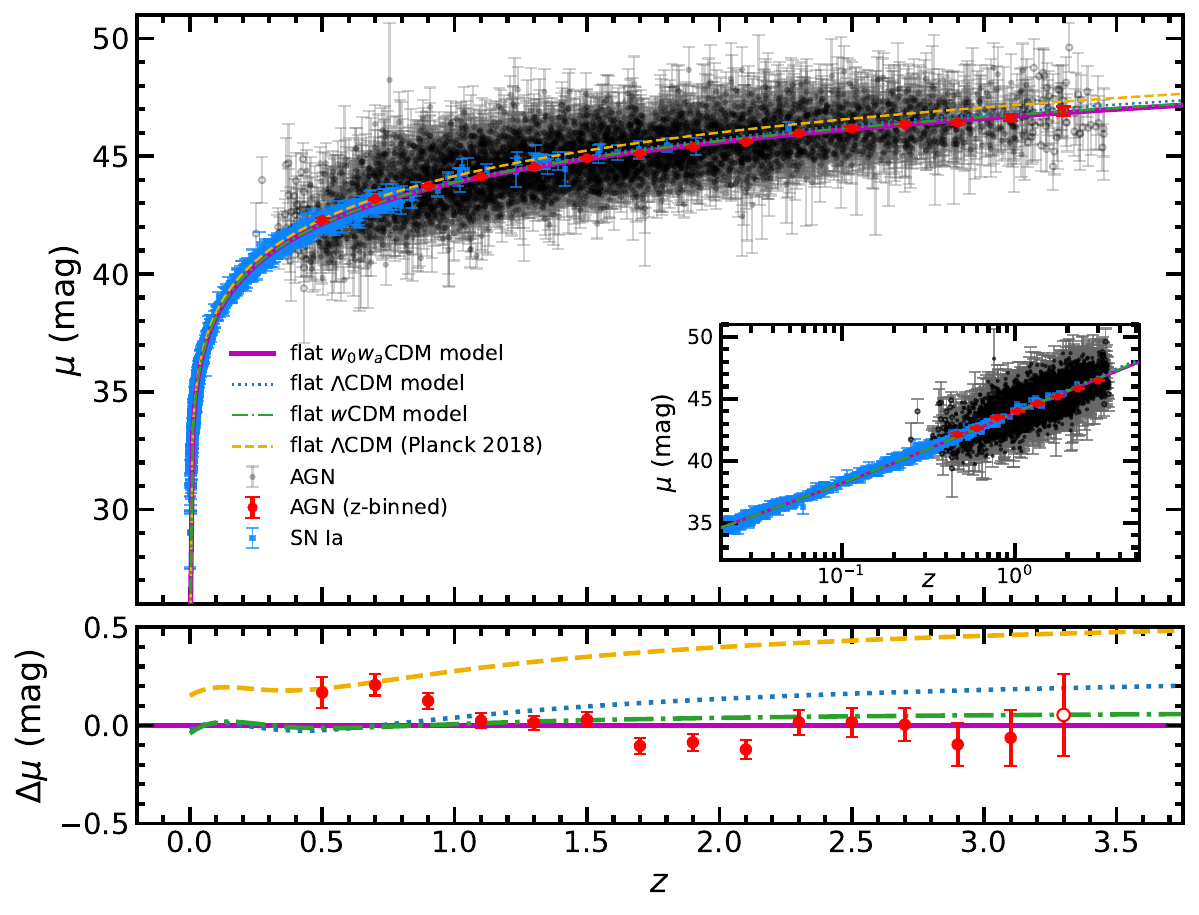}
\caption{Bias-corrected Hubble diagram for Pantheon+ SN Ia sample (blue circle symbols) and our plotted AGN sample (black circle symbols; this work). The filled circles correspond to our fiducial fitting AGN sample, where rest-frame $2500$ \AA\ is within the wavelength coverage of the spectrum. The open circles are sources where rest-frame $2500$ \AA\ is beyond the wavelength coverage of the spectrum, requiring an extrapolation of the AGN continuum. Open circles are not fitted, but included on the figure for completeness. The flat~$w_0w_a$CDM model fitted to the combined SN Ia + fiducial AGN sample is shown in magenta. The concordance $\Lambda$CDM and flat~$w$CDM models are shown for comparison as the blue dotted and green dash–dotted lines, respectively. The Planck 2018 result (\citep{PlanckCollaboration2020}; CMB power spectra, CMB lensing reconstruction, and BAO) is shown as the dashed yellow line, highlighting the Hubble tension. We overplot the AGN binned means and standard error on the means in red to guide the eye. The binned points are not used in the fit. The inset panel shows the Hubble diagram with redshift plotted using a logarithmic scale. The lower panel shows the binned residuals and fitted models relative to the best-fit flat~$w$CDM model. \label{fig:hd}}
\end{figure}

To establish this new class of cosmic probes, we start with the parent sample of spectroscopically-confirmed optically-unobscured (Type 1) AGNs with homogeneously-measured properties from the Sloan Digital Sky Survey (SDSS) \citep{Wu2022}. For a subsample of $\sim30,000$ AGNs in the SDSS Stripe 82 region, we assembled optical light curves from SDSS, Pan-STARRS, and the Zwicky Transient Facility (ZTF), spanning the $ugri$ bands \citep{Yu2025}. The light curves are modeled using a Gaussian process \citep{Kelly2014} that accounts for correlations in time and rest-frame wavelength (Figure~\ref{fig:lc}). We use a model that captures the observed spectral power density features over frequency $f$, which transitions from $f^0$ to $f^{-2}$ on characteristic ``damping'' timescales of hundreds of days, and displays additional features on shorter timescales (see Methods for details). Finally, we infer the rest-frame UV damping timescale $\tau_{\textrm{UV, RF}}$ and the amplitude $\sigma_{\textrm{UV}}$ for each AGN. From our fiducial fitting sample of $\resultNumAGNFitted$ AGNs with robustly measured variability, minimal galactic dust reddening ($E(B{-}V) < 0.05$), negligible host galaxy contamination, that lie between the redshift range $0.44 < z < 3.16$, we find that the AGN UV continuum luminosity scales with the variability parameters as,
\begin{equation}
    L_{2500\,\text{\AA}} = \resultLIntercept\, \left(\frac{\sigma_{\textrm{UV}} }{ \resultSigmaUVPivot \textrm{\, mag} } \right)^{\resultAlphaAGNL}\, \left(\frac{\tau_{\textrm{UV, RF}} }{ \resultTauUVRFPivot \textrm{\, days} } \right)^{\resultBetaAGNL},
\end{equation}
where $L_{2500\,\text{\AA}}$ is the monochromatic luminosity at 2500 \AA\ in the rest-frame, $\sigma_{\textrm{UV}}$, $\tau_{\textrm{UV, RF}}$ are the light curve characteristic variability amplitude and timescale. This yields an intrinsic scatter of $\resultScatterL$ in luminosity or $\resultScatterHD$ on the Hubble diagram. Our relation, jointly fitted with cosmological parameters as we describe in the paragraph below, is consistent with the relation derived by \cite{Suberlak2021} to $\lesssim 1\sigma$ (Figure~\ref{fig:l2500}). This correlation holds in narrow bins of redshift, demonstrating that its origin is neither due to selection effects nor shows significant redshift variations (see Methods for details). Previous work using the non-linear correlation between X-ray and UV flux has a comparable intrinsic scatter of $\sim 0.2 - 0.4$ dex in luminosity, depending on the sample definition \cite{Risaliti2015,Lusso2016,Lusso2017}. However, the luminosity self-correlation results in a much larger dispersion on the Hubble diagram of $\sim 1.4$ mag \citep{Lusso2025} compared to ours.

%\emph{The Hubble diagram and cosmo inference. Cosmo results.}
Similar to what is done with SN Ia as standard candles (e.g., \cite{Tripp1998,Kessler2017}), we fit the AGN distance modulus, $\mu^{\rm{AGN}}$, using the relation,
\begin{equation}
    \mu^{\rm{AGN}} = m_{2500\,\text{\AA}} - M_{0} - \alpha\, \log \sigma_{{\rm{UV}}} - \beta\, \log \tau_{{\mathrm{UV, RF}}}, 
\label{eq:mumodel}
\end{equation}
where $m_{2500\,\text{\AA}}$ is the rest-frame $2500$ \AA\ apparent magnitude of the accretion disk continuum calculated from spectral decomposition of SDSS data; $M_0$ is the fiducial absolute magnitude; and $\alpha$, $\beta$ are the regressor coefficients that characterize the light curve. Using Bayesian inference, we jointly fit the SN Ia and cosmological parameters for three dark energy models, all assuming a spatially flat Universe: (i) flat $\Lambda$CDM, where $\Lambda$ is the cosmological constant; (ii) flat $w$CDM, in which the dark energy equation-of-state parameter $w$ is constant; and (iii) flat $w_0w_a$CDM, where the equation-of-state follows the Chevallier–Polarski–Linder (CPL) parameterization $w(a) = w_0 + w_a(1 - a)$, allowing for redshift evolution through the parameter $w_a$ \citep{Chevallier2001,Linder2003}. We fit each cosmological model to two data combinations. First, we use SN~Ia + SH0ES data from the Pantheon+ compilation \citep{Brout2022}. Second, we augment this SN~Ia dataset with our newly derived AGN Hubble diagram, extending the redshift baseline to $z\sim3.5$ and providing complementary constraints on the expansion history. Our AGN Hubble diagram and residual model comparison are shown in Figure~\ref{fig:hd}. This figure demonstrates that the addition of AGN at higher redshift on the Hubble diagram favors a cosmology with evolving dark energy.

\begin{figure}
\centering
\includegraphics[width=1\textwidth]{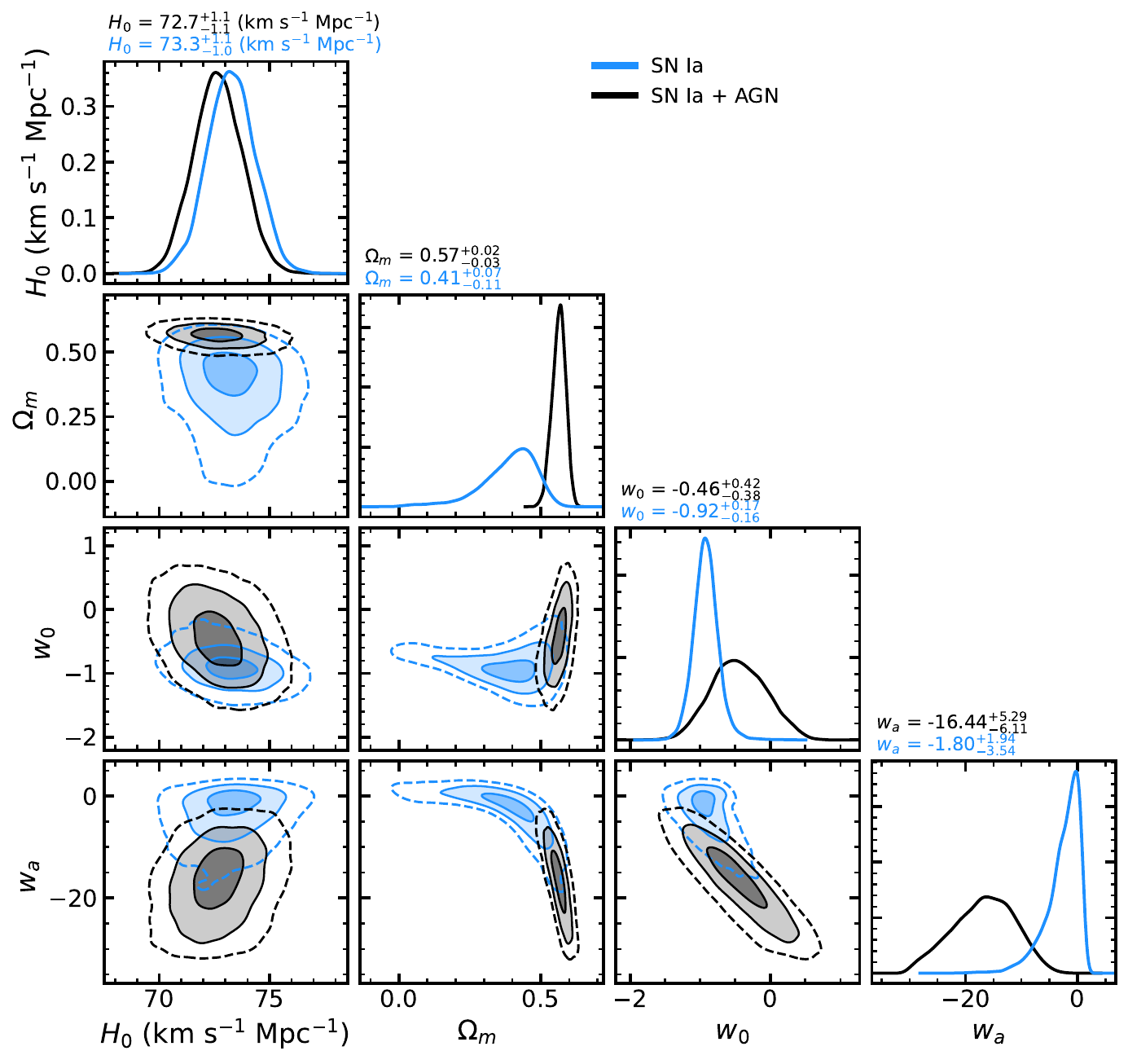}
\caption{Posterior distribution for flat~$w_0w_a$CDM cosmological parameters for the Pantheon+ SN Ia \citep{Brout2022} (shown in light blue) and combined SN Ia + AGN fiducial fitting sample (this work, shown in gray). The posterior distributions for each parameter are shown along the diagonal. The text in the upper right gives the posterior medians and 16th and 84th percentiles (1$\sigma$ credible intervals). The 2D contour plots show the covariances between each pair of parameters. The darker and lighter shaded regions are the 1$\sigma$ and 2$\sigma$ contours, and dashed curves delimit the 3$\sigma$ contours. 
\label{fig:cornerw0waCDM}}
\end{figure}

Figure~\ref{fig:cornerw0waCDM} shows the resulting marginalized posterior distributions for the cosmological parameters of the flat~$w_0w_a$CDM. With the inclusion of AGN data, we find that an evolving dark energy cosmology (flat~$w_0w_a$CDM) is preferred/favored over the flat~$w$CDM model at the $\resultzonecutSigmaFlatwCDM{-}\resultSigmaFlatwCDM\sigma$ significance level and over the flat~$\Lambda$CDM model at the $\resultzonecutSigmaFlatLambdaCDM{-}\resultSigmaFlatLambdaCDM\sigma$ significance level using the Bayesian evidence ratio. The range in model significance depends on the redshift range adopted. We stress that this finding is contingent on the fact that we have fully characterized all the sources of systemic error for variable AGN. Our results from this completely different and independent cosmic probe lends additional support to the $2.8-4.2\sigma$ evidence for the CPL model reported by the DESI collaboration using BAO \citep{DESICollaboration2025}. The age of the Universe from our preferred model is $\resultAgeUniverse$. Similar to the cosmological inference for the CPL model using higher-redshift SN Ia from the Dark Energy Survey \citep{DESCollaboration2024}, our results are also in tension with the ages of the oldest observed globular clusters \citep{Valcin2025,Ying2025,CimattiMoresco2023} and the determination of $\Omega_m$ from the Planck CMB data. To alleviate these early and late time tensions, alternative proposals for evolving  dark energy models are actively being explored \citep{Poulin2023Review}.

Continued investigation into the universality and intrinsic scatter of the AGN variability standardization can improve upon our results, and is a critical next step in establishing AGN variability as a mature cosmological probe. In particular, the advent of high-quality, high-cadence AGN light curves  that will soon become available from the Vera C. Rubin Observatory \citep{Ivezic2019} will allow for more robust tests of redshift-dependent biases, selection effects, and evolutionary systematics. At the same time, realizing the full potential of AGN cosmography requires progress on the calibration of AGN distances independent of the SN Ia distance ladder. Several promising avenues already exist. For example, a small sample of nearby AGNs with Cepheid distances already exists (e.g., \citep{Yuan2020,Yuan2021}), providing an absolute distance scale that can anchor AGN variability relations without relying on SN Ia. At farther distances, time-delay cosmography from strongly lensed quasars is another approach \citep{Birrer2024}. Time delays between the UV continuum and broad line variability (reverberation mapping) correlate with AGN luminosity (e.g., \citep{Watson2011,Martinez-Aldama2019}). Infrared reverberation mapping yields a geometric dust-sublimation radius that has been shown to correlate tightly with luminosity, offering another route to distance calibration (e.g., \citep{Yoshii2014,Yang2020}). Nevertheless, it is clear that AGNs can provide a complementary probe of the nature of dark energy and the expansion history of the Universe.

%TC:ignore
\section{Methods}

\subsection{Data}

Our parent sample comprises Type 1 AGNs confirmed by spectroscopy with well-measured properties from Sloan Digital Sky Survey (SDSS) spectroscopy from \cite{Wu2022}. For a subsample of $\sim\:$30,000 AGNs located in the SDSS Stripe 82 region, we collected their optical light curves from SDSS, Pan-STARRS, and the Zwicky Transient Facility (ZTF) \citep{Yu2025}. SDSS light-curves are available in the $u$, $g$, $r$, $i$, and $z$ bands; from Pan-STARRS in the $g$, $r$, $i$, and $z$ bands; and from the ZTF in the $g$, $r$, and $i$ bands. We omit data from the $z$ band, which tends to be nosier and suffer from the strongest host galaxy contamination, especially at lower redshifts. We note that photometry from different surveys is calibrated to a common photometric standard using spectrophotometry.  We recalibrate the photometric uncertainty of each survey individually using standard stars observed in the same sky region as our AGNs. The resulting combined light curves have a baseline of $\gtrsim\:$20 years, with an average of approximately 600 photometric data points (across all bands) and a median cadence of around 4 days. During the 20-year baseline, more than 98\% of our AGNs show variability exceeding 0.5 magnitude in the $r$ band.

We apply a set of quality cuts to construct the final AGN sample prior to light-curve modeling. We require SDSS spectra with a minimum length of 1700 days, a median per-pixel signal-to-noise ratio $\geq 3$ and Galactic reddening $E(B-V) \leq 0.05$. To ensure intrinsic variability is detected with confidence, we retain only objects with a $g$-band light curve variability significance of $\chi_g^2 \geq 10$ \citep[e.g.,][]{Butler2011}, where $\chi^2$ is the reduced Chi squared statistic representing the deviation of the observed light curve from a constant-flux model, thus excluding sources consistent with statistical noise. To avoid underestimating variability amplitudes in bands affected by intergalactic Ly$\alpha$ absorption, we remove measurements in the $u$ band for $z \gtrsim 1.3$, the $g$ band for $z \gtrsim 1.9$, the $r$ band for $z \gtrsim 3.1$, and the $i$ band for $z \gtrsim 4.1$. We further exclude objects that exhibit measurable host-galaxy contribution, Balmer continuum or iron pseudo-continuum from the BLR that would otherwise contaminate the AGN continuum.

Our sample now consists of $\resultNumAGNPlotted$ AGNs (\textit{plotted sample}). We then cut this list to AGN between $0.44 < z < 3.16$, ensuring the rest-frame 2500~\AA\ reference wavelength is within the SDSS spectral wavelength coverage rather than being extrapolated, resulting in $\resultNumAGNFitted$ AGNs (\textit{fiducial fitting sample}). Finally, $z\lesssim1$ AGN are significantly more affected by systematics such as host galaxy contamination and source extendedness that makes our light curves and AGN apparent magnitudes less reliable. Therefore, we further restrict the fitting to $z>1$, resulting in $\resultzonecutNumAGNFitted$ AGNs (\textit{restricted fitting sample}). Using this restricted fitting sample, we find the cosmological parameters are unchanged within $1\sigma$, but it does modestly decrease the CPL model significance, because fewer points are included. Therefore, we quote the significance using the full $0.44 < z < 3.16$ and restricted $1.0 < z < 3.16$ ranges throughout this work. 

\subsection{Light curve fitting}

An active galactic nucleus (AGN) is powered by accretion onto a supermassive black hole. Closest in ($\lesssim 0.01$ pc) lies the accretion disk, a geometrically thin, optically thick flow that radiates strongly in the UV/optical. Above it sits the X-ray corona, a hot ($\sim 10^9$ K) plasma that Compton-upscatters disk photons into the X-ray band. On scales of $\sim 0.01–0.1$ pc the broad-line region (BLR) consists of dense, fast-moving clouds producing Doppler-broadened emission lines that echo the disk variability. Surrounding these inner regions at $\sim 1–10$ pc is the dusty torus, a clumpy molecular structure that absorbs and re-emits radiation in the infrared, whose orientation largely determines whether we observe the AGN as Type 1 (optically unobscured) or Type 2 (optically obscured) \citep{Urry1995}.

We model the optical variability of broad-band AGN light curves as a linear combination of two components: intrinsic variations from the accretion disk UV continuum (possibly with an X-ray reprocessing component), and the broad-line region (BLR). We model the accretion disk variations using a Gaussian process (GP) with a summation of two damped random walk (DRW) kernels. This kernel has the form of an overdamped simple harmonic oscillator, which is a type of second-order continuous auto-regressive moving average (CARMA) process. This model captures the known deviations from a first-order CARMA process (DRW) on short timescales [e.g., \cite{Mushotzky2011,Kasliwal2015,Smith2018,Stone2022,Yu2022,Yu2025}].

Each band $b$ is associated with a characteristic damping timescale $\tau_b$, and the joint latent process across all bands has a covariance,
\begin{equation}
k^{\rm{lat}}_{b_1 b_2}(\Delta t)
= \frac{2\sqrt{\tau_{b_1}\tau_{b_2}}}{\tau_{b_1}+\tau_{b_2}}
\kappa \exp\!\left[-\frac{|\Delta t|}{\bar{\tau}_{b_1 b_2}}\right] + \frac{2\sqrt{\tau^{\rm{fast}}_{b_1}\tau^{\rm{fast}}_{b_2}}}{\tau^{\rm{fast}}_{b_1}+\tau^{\rm{fast}}_{b_2}}
(1-\kappa) \exp\!\left[-\frac{|\Delta t|}{\bar{\tau}^{\rm{fast}}_{b_1 b_2}}\right],
\end{equation}
where $\bar{\tau}_{b_1 b_2}$, ($\bar{\tau}^{\rm{fast}}_{b_1 b_2}$) is the harmonic mean damping (fast) timescale between bands $b_1$ and $b_2$, and $\kappa$ sets the relative amplitude between the second and first order terms. This form is symmetric and positive semi-definite (PSD). The observed flux in each band is modeled as a linear combination of a continuum $y^{\rm{cont}}$ and a time-delayed continuum component:
\begin{equation}
y_b(t)
= \sigma_b^{\mathrm{cont}}\,y^{\rm{cont}}(t - \ell_b^{\mathrm{disk}})
+ \sigma_b^{\mathrm{BLR}}\,y^{\rm{cont}}(t - \ell_b^{\mathrm{disk}} - \ell_b^{\mathrm{BLR}}),
\end{equation}
where $\sigma_b^{\mathrm{cont}}$ and $\sigma_b^{\mathrm{BLR}}$ are the amplitudes of the continuum and BLR in each band $b$ respectively, 
$\ell_b^{\mathrm{disk}}$ is the wavelength-dependent disk lag between bands, and
$\ell_b^{\mathrm{BLR}}$ is the additional delay from the response from the broad-line region. The resulting covariance between two observations $(t_1,b_1)$ and $(t_2,b_2)$ is:
\begin{align}
k \big((t_1,b_1),(t_2,b_2)\big)
&= \sigma_{b_1}^{\mathrm{cont}}\sigma_{b_2}^{\mathrm{cont}}\,
k^{\rm{lat}}_{b_1 b_2}\!\big(t_1-\ell_{b_1}^{\mathrm{disk}},\,t_2-\ell_{b_2}^{\mathrm{disk}}\big) \nonumber\\
&\quad + \sigma_{b_1}^{\mathrm{cont}}\sigma_{b_2}^{\mathrm{BLR}}\,
k^{\rm{lat}}_{b_1 b_2}\!\big(t_1-\ell_{b_1}^{\mathrm{disk}},\,t_2-\ell_{b_2}^{\mathrm{disk}}-\ell_{b_2}^{\mathrm{BLR}}\big) \nonumber\\
&\quad + \sigma_{b_1}^{\mathrm{BLR}}\sigma_{b_2}^{\mathrm{cont}}\,
k^{\rm{lat}}_{b_1 b_2}\!\big(t_1-\ell_{b_1}^{\mathrm{disk}}-\ell_{b_1}^{\mathrm{BLR}},\,t_2-\ell_{b_2}^{\mathrm{disk}}\big) \nonumber\\
&\quad + \sigma_{b_1}^{\mathrm{BLR}}\sigma_{b_2}^{\mathrm{BLR}}\,
k^{\rm{lat}}_{b_1 b_2}\!\big(t_1-\ell_{b_1}^{\mathrm{disk}}-\ell_{b_1}^{\mathrm{BLR}},\,t_2-\ell_{b_2}^{\mathrm{disk}}-\ell_{b_2}^{\mathrm{BLR}}\big),
\label{eq:covariance}
\end{align}
where the first term represents the covariance between the continuum flux in both bands. The second and third terms represent the cross-covariances between the continuum in one band and the BLR component in the other, incorporating the appropriate BLR lag $\ell^{\rm{BLR}}_b$. The fourth term represents the covariance between BLR components in both bands, with both time arguments shifted by their respective lags.

%AGNs exhibit deviations from the DRW on short timescales. This behavior is typically attributed to variations in the strength of X-ray reprocessing with luminosity or smearing effects. The former explanation is consistent with models where intrinsic disc fluctuations drive the variability and propagate inward (e.g., \cite{Neustadt2022,Stone2023}) or models where the disk and corona are magnetically coupled \citep{Sun2020}. In these models, the X-ray reprocessing component is stronger when the SED is UV-dominated, as observed in the AGN Fairall 9 \cite{Hagen2024}. To model this effect, we introduce an additional component that modifies the continuum--continuum block of the covariance:

To account for disk lags, we also shift the times in each band with filter central wavelength $\lambda_b$ according to,
\begin{equation}
t = t_0 - \ell^{\rm{disk}} \left[\left(\frac{\lambda_b}{\lambda_0}\right)^{\beta^{\rm{disk}}} - 1\right],
\end{equation}
where $t^0$ is the original time, $\ell^{\rm{disk}}$ is the disk lag intercept at wavelength $\lambda_0$, and $\beta^{\rm{disk}}$ is the power law index (e.g., \citep{Cackett2007}). 

%Our model differs from the model of \citet{Zu2011} in several key ways. First, \citet{Zu2011} assume that the light curves in each band are convolutions of the driving shortest-wavelength light curve, which is assumed to be a DRW. In contrast, we assume a latent driving UV light curve that is not directly observed. Rather, each observed light curve is a convolution of the latent driving UV light curve (assumed to be a DRW) that is smeared over the filter bandpass. This convolved kernel has very similar properties to a damped harmonic oscillator or CARMA(2,1) process, following $f^{-2}$, but steepening on short timescales set by $w$ \citep{Tachibana2020}. The resulting power spectrum (PSD) of this kernel is well matched to observed PSDs from previous work \citep{Stone2022}. 
% This naturally captures the population-level bluer-when-brighter phenomenon, in which their optical/UV continuum spectrum becomes harder (bluer) during brighter phases (e.g., \cite{Giveon1999,Sakata2010,Ruan2014,Meusinger2011,Schmidt2012}). 

The ensemble variability amplitude and timescale in each band scales with rest-frame wavelength, $\lambda_{\rm{RF}}$, according to a power law \cite{Kelly2009,MacLeod2010,Suberlak2021,Stone2022}, possibly with a break near  $\lambda_{\rm{RF}} = 2500$ \AA\ \citep{Yu2022,Yu2025}. To capture these possibilities, the DRW timescale in each band for a given AGN is empirically described as,
\begin{align}
\tau_b(\lambda_{b,\mathrm{RF}}) = \tau_{\mathrm{UV}} \left( \frac{\lambda_{b,\mathrm{RF}}}{\lambda_{\mathrm{UV}}} \right)^{\eta_\tau},
\qquad
\sigma_b^{\mathrm{cont}}(\lambda_{b,\mathrm{RF}}) = \sigma_{\mathrm{UV}} \left( \frac{\lambda_{b,\mathrm{RF}}}{\lambda_{\mathrm{UV}}} \right)^{\eta_\sigma},
\end{align}
where $\lambda_{b,\mathrm{RF}}$ is the reference wavelength, and $\eta_\tau$ and $\eta_\sigma$ are the power-law indices that govern the wavelength dependence (both free parameters). The final rest-frame UV parameters used to build the luminosity relation are,
\begin{equation}
\begin{aligned}
    \tau_{\mathrm{UV,\, RF}} &\equiv \tau_{b}(2500\,\text{\AA})/(1 + z) \\
    \sigma_{\mathrm{UV}} &\equiv \sigma_{b}(2500\,\text{\AA}).
\end{aligned}
\end{equation}
The likelihood for each AGN with parameters $\theta$ (Table~\ref{tab:agn_fits_table}) is,
\begin{equation}
%\mathcal{L}(\theta) \equiv 
p(y | \boldsymbol{\theta}) =
\frac{1}{\sqrt{2\pi |\mathbf{K}|}} 
\exp\left( 
    -\frac{1}{2} (\mathbf{y} - \mathbf{m})^\top \mathbf{K}^{-1} (\mathbf{y} - \mathbf{m}) 
\right),
\end{equation}
where $\mathbf{K}$ is the covariance matrix with elements, 
\begin{equation}
K_{nm} = k(t_n, t_m) + \delta_{nm}\, \sigma_{{\rm diag}, n}^2,
\end{equation}
where $k$ is the covariance function (kernel) given by Equation~\ref{eq:covariance}, \( \delta_{nm} \) is the Kronecker delta and $\sigma_{{\rm diag}, n}^2$ is the variance of observation $y_n$ in magnitudes. The total variance $\sigma_{{\rm diag}, n}^2 = \sigma_{\rm{jit}}^2 + \sigma_{y, n}^2$, where $\sigma_{y, n}$ is the photometric error in magnitudes of point $n$ and $\sigma_{\rm{jit}}$ is an excess scatter (``jitter'') error term to model under-estimated photometric uncertainties. Physically, the observed flux is the sum of continuum and BLR fluxes, which would make magnitudes nonlinearly related to these components. However, since variability amplitudes are typically at the 0.1 to 0.5 mag level, we approximate the fluctuations as linear in magnitude space. In this approximation, the continuum and BLR contributions combine additively in magnitudes, and the GP remains Gaussian. We adopt a mean function $\mathbf{m}$ for long timescale linear detrending with elements:
\begin{equation}\label{eqn:mean_detrending}
    m_{b,n} = m_{b,0} + m_{1}\ (t_n - t_0) .
\end{equation}
where $m_b,0$ is the mean in each band, $m_1$ is a linear slope (shared across all bands), and $t_0$ is the mid-range of time. The motivation for this mean function detrending is to remove any very long-time-scale nonstationary component, such as from changes in the accretion rate over decades. Failure to model non-stationary trends can result in a bias in the inferred parameters in the stationary kernel \cite{Stone2022}. Our mean function is fitted simultaneously with the kernel to capture degeneracies between the kernel and the long-timescale nonstationary process (see Appendix~\ref{appx:length}).

Our model is implemented in the \textsc{EzTaoX} code \citep{Yu2025eztaox}, which uses \textsc{tinygp} to perform the light curve fitting \citep{Foreman-Mackey2024} and the Hamiltonian Monte Carlo No-U-Turn sampler \citep{Hoffman2011} with \textsc{numpyro} \citep{bingham2019pyro} to sample from the posterior probability distribution of each light curve batch. The model priors are given in Table~\ref{tab:multibandfitpriors}. By marginalizing over the nuisance parameters governing the wavelength scaling of the DRW parameters, deviations from DRW on short and long timescales, and BLR contamination for each AGN, we can infer the intrinsic parameters describing the variations in the rest-frame UV-emitting part of the accretion disk.

\begin{table}[!ht]
\centering
%\small

\begin{tabular}{l l}
\toprule
Parameter & Distribution \\
\midrule
$\eta_{\sigma}$ & $\mathcal{N}(-0.5,\,1)$ \\
$\eta_{\tau}$ & $\mathcal{N}(0.5,\,0.5)$ \\

$\ln(\tau_{\rm{UV}} / {\rm days})$ & $\mathcal{TN}\!\big(\ln(10^{2.5}(1+z)),\,1.2\,\ln(10.0);0.0,\,\ln(10^4 (1+z))\big)$ \\
$\ln( \sigma_{\rm{UV}} / {\rm mag})$ & $\mathcal{N}(-0.6,\,1.0)$ \\

$m_{b, 0}$ (mag) & $\mathcal{N}(0.0,\,0.2)$ \\
 
$m_{1}$ (mag / std days) & $\mathcal{N}(0.0,\,0.1)$ \\

$\ell^{\mathrm{disk}}$ (days) & $\mathcal{TN}\!\big(10.0,\,5.0;\,0,\,\infty\big)$ \\

$\beta^{\mathrm{disk}}$ & $\mathcal{TN}\!\big(4/3,\,0.2;\,0,\,\infty\big)$ \\

$\kappa$ & $\mathcal{U}(0.0,\,1.0)$ \\

$\ln(\tau^{\mathrm{fast}} / {\rm days})$ & $\mathcal{TN}\!\big(\ln(10.0(1+z)),\, \ln 25.0;\,0.0,\, \ln(100.0(1+z))\big)$ \\

$\ln \Delta \sigma_{b, \mathrm{BLR}}$ & $\mathcal{U}(-1,\,1)$ \\

$\ln(\ell^{\mathrm{BLR}}_b / {\rm days})$ & $\mathcal{U}\!\big(\ln 2.0,\,\ln 5000.0\big)$ \\

$\ln \sigma_{\rm{jit}}$ & $\mathcal{N}\!\big(\ln\bar{\sigma}_y,\,1.0\big)$ \\
\addlinespace[0.5em]
\bottomrule
\end{tabular}

\label{tab:multibandfitpriors}
\caption{Priors adopted in the AGN light-curve modeling. $\mathcal{U}(a,b)$ denotes a uniform prior over the range $[a,b]$, $\mathcal{N}(\mu,\sigma)$ a Normal prior with mean $\mu$ and standard deviation $\sigma$, and $\mathcal{TN}(\mu,\sigma; a,b)$ a truncated Normal prior restricted to the interval $[a,b]$. Note: $ \sigma_{b, \rm{BLR} } = \Delta \sigma^{\mathrm{BLR}}_b\, \sigma^{\mathrm{cont}}_b$; $\bar{\sigma}_{y}$ is mean of the quoted light curve error.
}
\end{table}

\subsection{Apparent magnitude measurements}
\label{sec:apparent_mag_measurements}

We estimate the rest-frame UV apparent magnitudes $m_{2500\,\text{\AA}}$ from the AGN continuum following a two-step procedure that combines photometric light curve information with spectroscopic modeling. We obtained optical spectroscopy from SDSS data release 18 for every source in our sample. To minimize scatter arising from intrinsic AGN variability between the photometric and spectroscopic epochs, we first perform an absolute flux calibration of the spectrum using the light curve. First, we measure the synthetic photometry from the observed spectrum in each band using the \textsc{speclite} code \cite{Kirby2024}. We then rescale the spectrum such that its synthetic photometry matches the mean of the light curve inferred from our Gaussian process fitting. This relative flux calibration reduces apparent magnitude offsets from variation apertures vs PSF photometry that would otherwise arise from short-term variability and ensures a consistent absolute flux calibration between spectroscopy and photometry. We correct for source extendeness by fitting a redshift-binned mean of the PSF$-$fiber magnitude differences, and subtracting the synthetic magnitudes from the interpolated function. Remaining errors in absolute flux calibration relative to SN Ia magnitudes will be absorbed by the $M_{0}$ term in the Hubble diagram fitting.

We re-fit the calibrated SDSS spectra using the \textsc{PyQSOFit} package \citep{Guo2018,Shen2019}, following the methodology of \cite{Wu2022}. The model includes the AGN power-law continuum, Fe\,\textsc{ii} emission, Balmer continuum, and broad/narrow emission lines, and host galaxy decomposition. However, we make several changes to the methodology of \cite{Wu2022}. First, we include host galaxy decomposition using principle component analysis with host galaxy templates from \cite{Bruzual2003}. This mostly affects low-redshift ($z\lesssim 1$) AGNs with clear host galaxy features such as Balmer breaks. Second, rather than a single power law, we fit the AGN continuum using a smoothly broken power law with a break near $5000$ \AA\ \citep{VandenBerk2001}. Next, we change the internal reddening model from a pure additive polynomial to an SMC bar extinction law \citep{Gordon2003} that is less degenerate with the shape of the AGN continuum curve. Finally, rather than fit the continuum in pre-defined emission-line-free windows, we fit the continuum over the full wavelength range using a Huber loss function that naturally down-weights the emission lines. This reduces sensitivity to absorption features and windows near the edges of the spectra. 

We model each spectrum with and without host contribution and with and without Balmer continuum emission and select the best model using the minimum Bayesian information criterion (BIC) statistic. AGNs that prefer models with Balmer continuum or host galaxy contributions are removed before constructing the fiducial and restricted fitting sample. From the best-fit model, we extract the dereddened AGN power-law continuum component and evaluate it at rest-frame $2500$\,\AA\ to obtain the monochromatic flux density $f_{2500\,\text{\AA}}$. The apparent magnitude at rest-frame $2500$\,\AA\ is then computed in the AB system as
\begin{equation}
m_{2500\,\text{\AA}} = -2.5 \, \log \left( \frac{f_{2500\,\text{\AA}}}{3631\,\mathrm{Jy}} \right),
\end{equation} where $f_{2500\,\text{\AA}}$ is taken from the power-law continuum component of the fitted spectral model. This procedure yields $m_{2500\,\text{\AA}}$ values that are spectroscopically decomposed, minimizing contamination from emission lines and the host galaxy. We perform 50 bootstrap Monte Carlo re-samples to estimate the uncertainties on $m_{2500\,\text{\AA}}$.

\subsection{Cosmological inference}
%{\color{red} We investigate the following three cosmological models. We explore the parameter estimation for these three cosmological models: Model A, B, C... Specify very clearly the models we investigate.}

We use a Bayesian method to jointly infer the cosmological parameters and the AGN light curve -- luminosity relation. We follow the methodology of \cite{March2018}, which corrects for Malmquist bias and model uncertainty. The distance modulus is,
\begin{equation}
    \mu^{\rm{AGN}} = m_{2500\,\text{\AA}} - M_{0} - \alpha\, \log \sigma_{{\rm{UV}}} - \beta\, \log \tau_{{\mathrm{UV,\ RF}}}
\label{eq:mumodel2}
\end{equation}
where $m_{2500\,\text{\AA}}$ is the rest-frame UV apparent magnitude, $M_0$ is the fiducial absolute magnitude, $\alpha$, $\beta$ are regressor coefficients. The apparent magnitudes $m$ are calculated from SDSS spectral decomposition, while $\sigma_{{\rm{UV}}}$ and $\tau_{{\mathrm{UV}}}$ are derived from the light curve fitting. 

The uncertainty on the distance modulus is,
\begin{equation}
\begin{aligned}
{\sigma_{\mu}^{\text{AGN}}}^2 &= 
    \sigma_{m}^2 +
    (\alpha\, \sigma_{\log \sigma_{\mathrm{UV}}})^2 +
    (\beta\, \sigma_{\log \tau_{\textrm{UV,\,RF}}})^2 \\ 
     & + 2\, \alpha\, \beta\, \mathrm{cov}({\log \sigma_{\mathrm{UV}}},\, \log \tau_{\textrm{UV,\,RF}}) + 
    \sigma_{z}^2 +
    \sigma_{\mathrm{lens}}^2 +
    \sigma_{s}^2
\end{aligned}
\end{equation}
where $\sigma_m$ is the uncertainty in apparent magnitude, $\sigma_{\log \sigma_{\text{UV}}}$ is the uncertainty in the rest-frame UV variability amplitude, $\sigma_{\log \sigma_{\text{RF}}}$ is the uncertainty in the rest-frame UV damping timescale, $\sigma_z$ is the uncertainty on the redshift, $\sigma_{\rm{lens}}$ is the uncertainty from lensing, and $\sigma_s$ is the intrinsic scatter in the relation. We adopt $\sigma_{\rm lens}(z) = \left(0.060\pm0.017\right)\, \!\left[d_{C}(z)/d_{C}(z{=}1)\right]^{3/2}$, where $d_{C}(z)$ is the comoving distance, which extrapolates to higher redshift more accurately than the prescription from \cite{Jonsson2010}. The $\mathrm{cov}({\log \sigma_{\mathrm{UV}}},\, \log \tau_{\textrm{UV, RF}})$ term is the covariance between the UV amplitude and damping timescale. We ignore the small covariance between other parameters.

The degeneracy between $H_0$ and $M_0$ requires us to calibrate our results to sources farther down the distance ladder. We use the Cepheid-calibrated SN Ia measurements from the Pantheon+ analysis \citep{Brout2022} for this purpose. The SN Ia overlap with our sample at redshifts of $z \sim 0.25 - 1.5$. The distance moduli of the combined sample are,
\begin{equation}
\mu_i = 
\begin{cases} 
      \mu_i^{\rm{AGN}} & i \in {\rm{AGNs}} \\
      \mu_i^{\rm{SN~Ia}} & i \in {\rm{SN\ Ia}}.
\end{cases}
\end{equation}
We calculate $\mu_i^{\rm{SN~Ia}}$ following Equation 1 of \cite{Brout2022}, and we use the Cephied distances for calibrator sources. The covariance is,
\begin{equation}
\textbf{C}_{ij} = 
\begin{cases} 
      \delta_{ij}\, {(\sigma^{\rm{AGN}}_{\mu,i}})^2 & i \in {\rm{AGNs}} \\
      C_{ij}^{\rm{SN~Ia}} & i \in {\rm{SN\ Ia}},
\end{cases}
\end{equation}
where $C_{ij}^{\rm{SN~Ia}}$ is the total (statistical+systematic) covariance from \cite{Brout2022}.

The probability distribution of $\mu_i$ on the Hubble diagram is,
\begin{equation}
    p(\mu_i | z_i, \theta, \mathscr{C}) = \frac{1}{\sqrt{ 2\pi |\mathbf{C}
|}} \exp \bigg( -\frac{1}{2}\ \Delta \mathbf{D}^T\ \mathbf{C}^{-1}\ \Delta \mathbf{D}\ \bigg) 
\end{equation}
where $\Delta \mathbf{D}_i = \mu_i - \mu_{\rm{model}}(z_i, \mathscr{C})$ is the residual vector, $\mathbf{C}$ is the covariance matrix, $\theta$ are the light curve fitting parameters and $\mu_{\rm{model}}(z_i, \mathscr{C}) = 5 \log(d_L(z_i, \mathscr{C}) / 10\ {\rm{pc}})$ is the model distance modulus, which depends on the set of cosmological parameters, $\mathscr{C} \in \{H_0, \Omega_m, w_a, w_0\}$ (see \cite{Hogg1999}).
%The observed data likelihood is,
%\begin{equation}
%    p(\mu | m, z, C, \theta, N) = C^N_{n} \bigg[ \int_{m_{\rm{lim}}}^\infty p(\mu | m, z, C, \theta)\ dm \bigg]^{N-n}\prod_{i \in \mathcal{A}_{\rm obs}} p(\mu_i | m_i, z_i, C, \theta).
%\end{equation}
%where $C^N_n$ is the binomial coefficient with $n$ being the number of observed AGNs and $N$ being the (unknown) total number of unobserved AGNs, and $m_{\rm{lim}}$ is the limiting magnitude of our sample.
% https://arxiv.org/pdf/0805.2946 Equation 3
% Note N is a free parameter
% Can be marginalized out using Gelman's trick
The observed data likelihood is,
\begin{equation}
    p(\mu | m, z, \mathscr{C}, \theta, N) \propto \bigg[ \int p(I{=}1|m,z)\ p(\mu | m, z, \mathscr{C}, \theta)\ dm\ \bigg]^{-n}\prod_{i \in \mathcal{A}_{\rm obs}} p(\mu | m, z_i, \mathscr{C}, \theta),
    \label{eq:hubble_likelihood}
\end{equation}
where $n$ is the number of AGNs in the observed data set. The second term is the product for the observed data set. The first term is the product for the missing data set, marginalized over all possible values of the missing data and over the total number of observed and missing AGNs $N$ assuming a uniform prior in $\log N$ (e.g., \cite{gelman2013bayesian, Kelly2007}). This corrects for incompleteness, which is a function of redshift and apparent magnitude. The selection function $p(I{=}1|m,z)$ is the probability that a source with apparent magnitude $m$ and redshift $z$ is included in the sample. Our method for estimating the completeness function is described in Appendix~\ref{sec:completeness}. We assume $p(I{=}1|m,z)=1$ for the SN Ia, because their magnitudes have already been bias-corrected \citep{Popovic2021}. We use the \textsc{dynesty} nested sampler \citep{Speagle2020} to estimate the posterior probability density and Bayesian evidence $\mathcal{Z}$ using $d\log \mathcal{Z} = 0.01$.

We compute the bias-corrected magnitudes by estimating the difference between the completeness-weighted expectation value of the apparent magnitudes and the model-predicted magnitudes.
\begin{equation}
\Delta m (z) =
\frac{\int m \, p\!\left(\mu | \mu_{\mathrm{model}}\right)\,
p(I{=}1 | m, z)\, dm}
{\int p\!\left(\mu | \mu_{\mathrm{model}}\right)\,
p(I{=}1 | m, z)\, dm} - m_{\mathrm{model}},
\end{equation}
where $m_{\mathrm{model}} = M_0 - \mu_{\mathrm{model}}$. The bias-corrected apparent magnitudes are $m_{\rm{corr}} = \Delta m(z) -  m$ where $m=m_{2500\,\text{\AA}}$ are the observed apparent magnitudes. This corrects for the systematic shift introduced by the preferential detection of brighter objects. These bias-corrected magnitudes and distance moduli are only used for visualization purposes in the bias-corrected figures. We always use the observed magnitudes and the completeness function for inference of cosmological parameters through Equation~\ref{eq:hubble_likelihood}.

\subsection{Cosmological models}\label{sec:model_selection}

\begin{table}[!ht]
\centering
%\small
\begin{tabular}{ll}
\hline
Parameter & Prior Distribution \\
\hline

%\multicolumn{2}{l}{\textbf{Supernovae}} \\
$M^0_{\rm SN}$ & $\mathcal{U}(-20, -18)$ \\[8pt]

%\multicolumn{2}{l}{\textbf{AGN}} \\
$M^0_{\rm AGN}$ & $\mathcal{U}(-26, -18)$ \\
$\alpha_{\rm AGN}$ & $\mathcal{U}(-20, 20)$ \\
$\beta_{\rm AGN}$ & $\mathcal{U}(-20, 20)$ \\
$\log \sigma_s$ & $\mathcal{U}(-2.2, 1.3)$ \\[8pt]

%\multicolumn{2}{l}{\textbf{Flatw0waCDM}} \\
$H_0$ & $\mathcal{U}(60, 80)$ \\
$\Omega_{m}$ & $\mathcal{U}(0, 1)$ \\
$w_0$ & $\mathcal{U}(-3, 1)$ \\
$w_a$ & $\mathcal{U}(-30, 1)$ \\[8pt]
\hline
\end{tabular}
\caption{Prior distributions for the cosmological model fitting. The quantity $\mathcal{U}(a, b)$ indicates a uniform prior between the minimum value $a$ and maximum value $b$. \label{tab:hubblepriors}}
\end{table}

\begin{figure}
\centering
\includegraphics[width=0.8\textwidth]{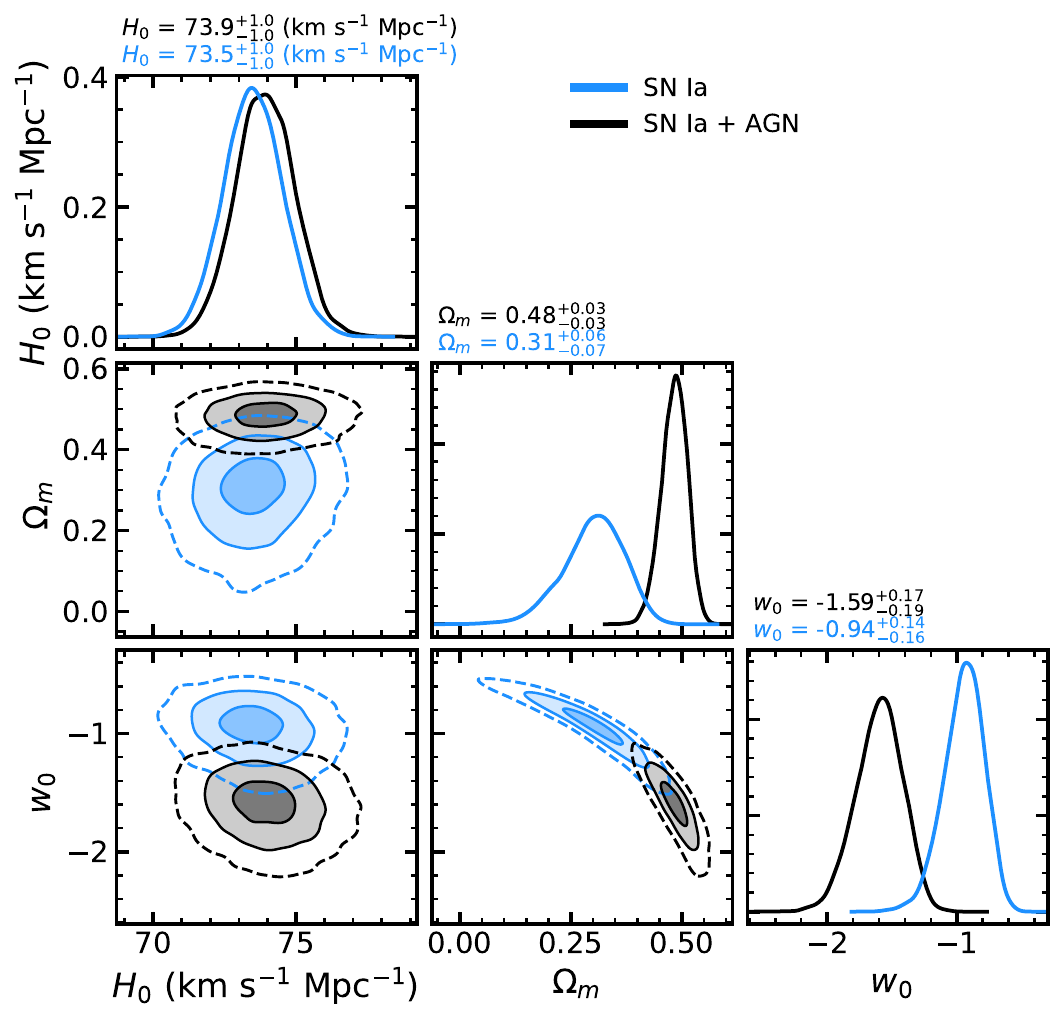}
\caption{Same as Figure~\ref{fig:cornerw0waCDM} but for the flat~$w$CDM model where $w$ is not fixed to -1. 
\label{fig:cornerwCDM}}
\end{figure}

\begin{figure}
\centering
\includegraphics[width=0.8\textwidth]{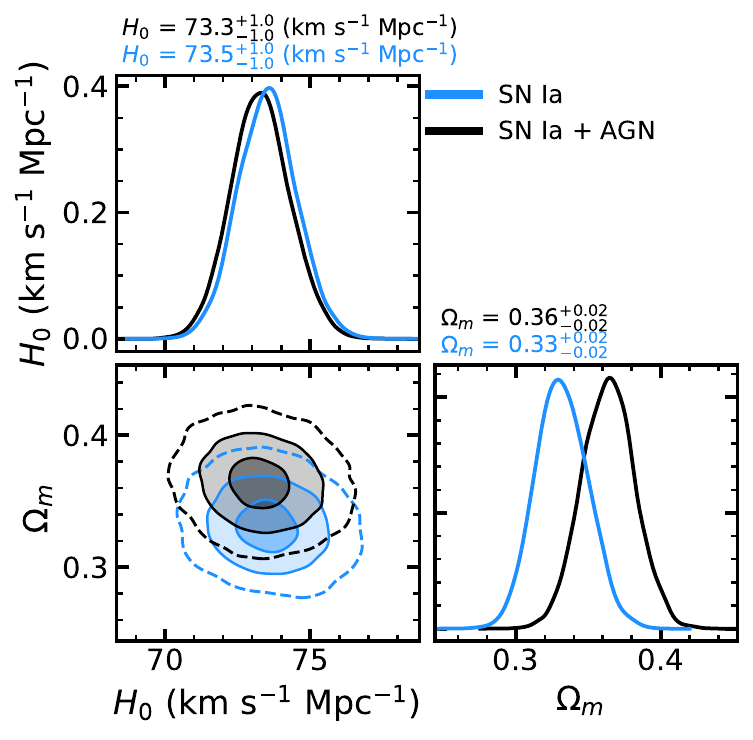}
\caption{Same as Figure~\ref{fig:cornerw0waCDM} but for the flat~$\Lambda$CDM model. 
\label{fig:cornerLCDM}}
\end{figure}

We perform Bayesian inference of cosmological parameters using the likelihood given by Equation~\ref{eq:hubble_likelihood} and uniform priors on all cosmological parameters (see Table~\ref{tab:hubblepriors}). We try three different dark energy models, all assuming a spatially flat Universe: 
\begin{enumerate}
    \item \textbf{flat $\Lambda$CDM}, in which dark energy is described by a cosmological constant with equation-of-state parameter fixed at $w=-1$;
    \item \textbf{flat $w$CDM}, in which the dark energy equation-of-state parameter $w$ is constant;
    \item \textbf{flat $w_0w_a$CDM}, where the dark energy equation-of-state follows the Chevallier–Polarski–Linder (CPL) parameterization $w(a) = w_0 + w_a(1 - a)$, allowing for redshift evolution.
\end{enumerate}

Each cosmological model is fit to two data combinations. First, we analyze the SN~Ia~+~SH0ES sample from the Pantheon+ compilation, which includes over 1,700 uniformly calibrated spectroscopically confirmed SNe~Ia to obtain precise distance estimates. The SH0ES Cepheid ladder anchors the absolute luminosity scale of the sample \citep{Riess2022}. Second, we perform a joint fit combining SN~Ia and AGN, extending the redshift baseline and providing complementary constraints on the expansion history.

As shown in Table~\ref{tab:cosmoparams}, our SN Ia only fits reproduce the results of \citet{Brout2022} to within $1\sigma$ across all models, confirming the robustness of our analysis pipeline. In the flat~$\Lambda$CDM and flat~$w$CDM cases, we obtain $H_0 = 73.53 \pm 1.01~\mathrm{km\,s^{-1}\,Mpc^{-1}}$ and $\Omega_m = 0.33 \pm 0.02$, fully consistent with the Pantheon+~\&~SH0ES values of $H_0 = 73.6 \pm 1.1$ and $\Omega_m = 0.334 \pm 0.018$. Likewise, our inferred dark energy parameters of $w_0 = -0.92 \pm 0.18$ and $w_a = -2.59 \pm 3.08$ are in excellent agreement with the Pantheon+~\&~SH0ES values of $w_0 = -0.93\, \pm\, 0.15$ and $w_a = -0.1^{+0.9}_{-2.0}$. Minor differences can be attributed to differences in the assumed priors and posterior sampling codes. Figure~\ref{fig:cornerLCDM}, Figure~\ref{fig:cornerwCDM} and Figure~\ref{fig:cornerw0waCDM} show the resulting marginalized posterior distributions for the cosmological parameters $\Omega_m$, $H_0$ (for flat~$\Lambda$CDM), $w$ (for flat~$w$CDM) and $w_0$, $w_a$ (for flat~$w_0w_a$CDM), respectively. The corresponding credible intervals and Bayesian evidences are summarized in Table~\ref{tab:cosmoparams}. We find that the inclusion of AGN data at higher redshift leads to a modestly larger matter density for all models compared to \citep{Brout2022}. Additionally, the inclusion of AGN data pushes the $w_a$ constraints to very negative values compared to lower redshift SN Ia only.

We compare cosmological models using the marginal likelihood, or Bayesian evidence, $\mathcal{Z}$, defined as the likelihood integrated over the prior volume. We kept the adopted priors (Table~\ref{tab:hubblepriors}) fixed across all models to ensure a fair comparison. Differences in the logarithm of the evidence, $\Delta \ln \mathcal{Z}$, provide a natural metric for relative model plausibility \citep{Jeffreys1983}. A positive $\Delta \ln \mathcal{Z}$ favors the model with larger evidence. For interpretability, $\Delta \ln \mathcal{Z}$ can be converted to a Gaussian equivalent significance,
\begin{equation}
\sigma_{\mathrm{eq}} = \sqrt{2\, \Delta\, \ln \mathcal{Z}},
\end{equation}
where $\sigma_{\mathrm{eq}}$ is the Gaussian equivalent significance \citep{Kass1995}.

We compute $\ln \mathcal{Z}$ for three cosmological scenarios. The dynamical dark energy model flat~$w_{0}w_{a}$CDM yields the highest evidence, $\ln \mathcal{Z}=\resultLogZFlatwZerowaCDM \pm \resultLogZerrFlatwZerowaCDM$. Relative to this preferred model, flat~$w$CDM is disfavored at $\Delta \ln \mathcal{Z}=\resultDeltaLogZFlatwZeroWaCDMFlatwCDM$, corresponding to $\resultSigmaFlatwZeroWaCDMFlatwCDM\sigma$ for AGNs in the range $0.44<z<3.16$ and $\resultzonecutSigmaFlatwZeroWaCDMFlatwCDM\sigma$ for the subset at $1<z<3.16$. The standard flat~$\Lambda$CDM model fares worse, with $\Delta \ln \mathcal{Z}=\resultDeltaLogZFlatwZeroWaCDMFlatLambdaCDM$, equivalent to $\resultSigmaFlatwZeroWaCDMFlatLambdaCDM\sigma$ over $0.44<z<3.16$ and $\resultzonecutSigmaFlatwZeroWaCDMFlatLambdaCDM\sigma$ for $1<z<3.16$. According to the Jeffreys' scale \citep{Jeffreys1983}, the flat~$w_{0}w_{a}$CDM model is decisively preferred over the flat~$w$CDM and flat~$\Lambda$CDM models. We therefore conclude that the data favor a time-varying dark energy equation of state over both constant-$w$ and cosmological-constant models.

\subsubsection{Cosmological Implications}
The most favored cosmological scenario in our analysis is one with an evolving dark energy equation-of-state. As shown in Sec.~\ref{sec:model_selection}, we find $\resultzonecutSigmaFlatwZeroWaCDMFlatwCDM{-}\resultSigmaFlatwZeroWaCDMFlatwCDM\sigma$ evidence in support of a ``thawing model'', a flat $w_0 w_a$CDM in which the equation-of-state parameter is highly negative at earlier epochs and increases with time, compared to flat $w$CDM. Our improved constraints on $w(z)$ are enabled by the extended redshift range probed by AGN. However, we caution that at high redshift, the distance moduli are most sensitive to the completeness corrections. Meanwhile, at low redshift, our data are most affected by other sources of error like host galaxy contamination. Nevertheless, an evolving dark energy would be in agreement with BAO results \citep{DESICollaboration2025} and reinforces hints from the DES-SN analysis \citep{Abbott2022}. 

Our results favor a cosmological model that deviates significantly from the standard $\Lambda$CDM framework. As shown in Table~\ref{tab:cosmoparams}, the inferred matter density $\Omega_m = \resultOmZero$, is substantially higher than both the Planck and DESI DR2 (DESI only) results. It remains consistent within $\sim 1.5 \sigma$ of the $\Omega_m = 0.495^{+0.033}_{-0.043}$ obtained by the DES-SN analysis \citep{Abbott2022} in their flat $w_0 w_a$CDM model, indicating that the Universe remained more strongly matter-dominated over cosmic time. This leads to enhanced deceleration and smaller luminosity distances compared to $\Lambda$CDM expectations. The dark energy equation-of-state is described by $w_0 = \resultwZero$ and $w_a =\resultwa$, with a present-day value $w_0$ that is less negative than a cosmological constant ($w=-1)$, leading to weaker late-time acceleration relative to $\Lambda$CDM. A strongly negative $w_a$ implies a rapid evolution toward $w(z) = w_a$ at early times, causing the dark energy density to decay faster than matter. Consequently, dark energy remains subdominant throughout most of the expansion history and only becomes dynamically relevant at recent epochs, never reaching the degree of dominance seen in $\Lambda$CDM.

%This parameter combination yields an expansion history in which matter governs nearly the entire timeline of the Universe, with the observed Hubble diagram reflecting a suppressed distance–redshift relation consistent with this scenario. 
Because our calibration is anchored to SN Ia, our inferred $H_0$ is set by the SN Ia distance scale. In our flat $w_0 w_a$CDM model, the enhanced matter content shortens the age of the Universe to $t_0 = \resultAgeUniverse$, substantially younger than the $\sim 13.4$~Gyr age for the $\Lambda$CDM Universe from Planck data \citep{PlanckCollaboration2020}. 
Such a short $t_0$ is in immediate tension with multiple stellar chronometers that independently point to a Universe older than $\gtrsim 12$~Gyr. The oldest Milky Way globular clusters date to $\gtrsim 12$--$13$~Gyr \citep{Valcin2025,Ying2025,CimattiMoresco2023}. White-dwarf cooling sequences reach $13.9\pm0.8$~Gyr with Gaia parallaxes \citep{Fouesneau2018}. Nearby ultra-faint galaxies quenched early, with average quenching times of $\gtrsim 12.8$~Gyr \citep{Durbin2025,Brown2014}. Extremely metal-poor halo stars, including the Methuselah subgiant HD~140283, likewise yield $\gtrsim 12$~Gyr \citep{Guillaume2024,Frebel2007}. Reconciling our $t_0=\resultAgeUniverse$ with this ensemble would require substantial systematics across these independent methods or nonstandard stellar physics.

\begin{table}
\centering
\setlength{\tabcolsep}{4pt} % compact spacing
\begin{tabular}{lccccc}
\toprule
Dataset & $H_0$\tnote{a} & $\Omega_m$ & $w_0$ & $w_a$ & $\ln \mathcal{Z}$ \\
\midrule
\multicolumn{6}{l}{\underline{\textbf{flat $\Lambda$CDM}}} \\
Pantheon+ \& SH0ES & $73.6 \pm 1.1$ & $0.334 \pm 0.018$ & $-1$ & -- & -- \\
DES-SN5YR & -- & $0.352 \pm 0.017$ & $-1$ & -- & -- \\
Planck 2018 & $67.66 \pm 0.42$ & $0.3111 \pm 0.0056$ & $-1$ & -- & -- \\
DESI DR2 & -- & $0.2975 \pm 0.0086$ & $-1$ & -- & -- \\
\textbf{SN~Ia} & \textbf{$73.53 \pm 1.01$} & \textbf{$0.33 \pm 0.02$} & \textbf{--} & \textbf{--} & \textbf{$886.6 \pm 0.1$} \\
\textbf{SN~Ia + AGN} & \textbf{$73.29 \pm 1.02$} & \textbf{$0.36 \pm 0.02$} & \textbf{--} & \textbf{--} & \textbf{$5449.1 \pm 0.2$} \\
\midrule
\multicolumn{6}{l}{\underline{\textbf{flat $w$CDM}}} \\
Pantheon+ \& SH0ES & $73.5 \pm 1.1$ & $0.309^{+0.063}_{-0.069}$ & $-0.90 \pm 0.14$ & -- & -- \\
DES-SN5YR & -- & $0.264^{+0.074}_{-0.096}$ & $-0.80^{+0.14}_{-0.16}$ & -- & -- \\
DESI DR2 & -- & $0.2969 \pm 0.0089$ & $-0.916 \pm 0.078$ & -- & -- \\
\textbf{SN~Ia} & \textbf{$73.51 \pm 1.02$} & \textbf{$0.30 \pm 0.07$} & \textbf{$-0.95 \pm 0.15$} & \textbf{--} & \textbf{$883.2 \pm 0.1$} \\
\textbf{SN~Ia + AGN} & \textbf{$73.93 \pm 1.03$} & \textbf{$0.48 \pm 0.03$} & \textbf{$-1.60 \pm 0.18$} & \textbf{--} & \textbf{$5453.1 \pm 0.2$} \\
\midrule
\multicolumn{6}{l}{\underline{\textbf{flat $w_0w_a$CDM}}} \\
Pantheon+ \& SH0ES & $73.3 \pm 1.1$ & $0.403^{+0.054}_{-0.098}$ & $-0.93 \pm 0.15$ & $-0.1^{+0.9}_{-2.0}$ & -- \\
DES-SN5YR & -- & $0.495^{+0.033}_{-0.043}$ & $-0.36^{+0.36}_{-0.30}$ & $-8.8^{+3.7}_{-4.5}$ & -- \\
DESI DR2 & -- & $0.352^{+0.041}_{-0.018}$ & $-0.48^{+0.35}_{-0.17}$ & $< -1.34$ & -- \\
\textbf{SN~Ia} & \textbf{$73.28 \pm 1.07$} & \textbf{$0.39 \pm 0.10$} & \textbf{$-0.92 \pm 0.18$} & \textbf{$-2.59 \pm 3.08$} & \textbf{$882.2 \pm 0.1$} \\
\textbf{SN~Ia + AGN} & \textbf{$72.66 \pm 1.08$} & \textbf{$0.56 \pm 0.02$} & \textbf{$-0.45 \pm 0.39$} & \textbf{$-16.73 \pm 5.44$} & \textbf{$5460.6 \pm 0.2$} \\
\bottomrule
\end{tabular}
\begin{tablenotes}
\item[a] Units: km s$^{-1}$ Mpc$^{-1}$.
\end{tablenotes}

\caption{Marginalized cosmological parameters and bayesian evidence for the three cosmological models tested in this work. We quote the median and 1 sigma uncertainty. We quote the SN Ia results (without external priors) found from Pantheon+ \& SH0ES \citep{Brout2022} and DES-SN5YR \citep{DESCollaboration2024}. We also show the results from Planck 2018 analysis with CMB power spectra, CMB lensing reconstruction, and BAO \citep{PlanckCollaboration2020}, and DESI DR2 (DESI only) \citep{DESICollaboration2025}. The last two columns under each model are our results for SN Ia and SN Ia + AGN.
\label{tab:cosmoparams}
}
\end{table}

\subsection{Standardizability}

\begin{figure}[!ht]
\centering
\includegraphics[width=1\textwidth]{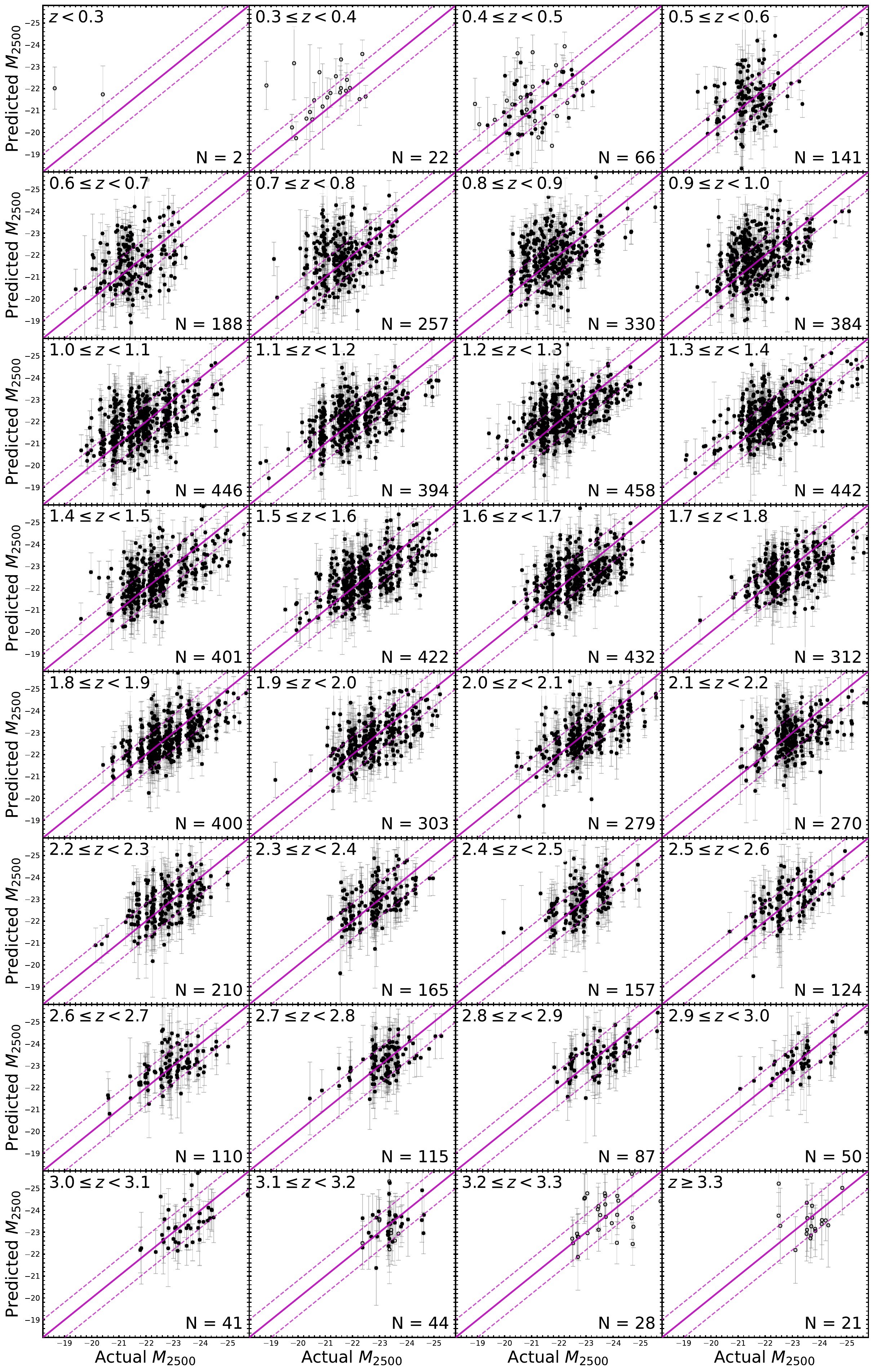}
\caption{Predicted absolute magnitude from Equation~\ref{eq:mumodel2} vs. actual, bias-corrected absolute magnitude $M_{2500\,\text{\AA}}$ in bins of redshift. The redshift ranges are shown in the upper left of each figure panel. The number of AGNs in the bin are shown in the lower right of each panel. The solid magenta lines show the $y=x$ prediction. The dashed magenta lines are the $\pm 1\sigma$ intrinsic scatter bounds. The predicted absolute magnitudes track closely with the actual absolute magnitudes regardless of redshift, indicating that our results are not due to a selection bias or redshift dependence. The filled circles indicate AGNs where rest-frame $2500$ \AA\ is within the wavelength coverage of the spectrum. The open circles are sources where rest-frame $2500$ \AA\ is beyond the wavelength coverage of the spectrum, requiring an extrapolation of the AGN continuum. 
% The colors from red to blue denote relative error $\sigma_{M_{2500}}/M_{2500}$ less than 0.02, 0.02-0.04, 0.04-0.06, 0.06-0.08, and greater than 0.08, respectively.
\label{fig:predicted_Mi_binned}
}
\end{figure}

Distances to standardizable candles are vulnerable to the \emph{circularity problem} \citep{Petrosian2022}. The circularity problem refers to the degeneracy between cosmology and the correlation parameters used to calculate the distances. To determine whether AGN variability is a good distance indicator, we must establish that there is no redshift dependence in the regression independent of cosmology and selection effects. We must also ensure that the correlation is not purely due to selection biases. For example, only the most luminous AGNs are likely to be detected at increasing distances. 

\subsubsection{Selection biases}

To test whether the correlation in our sample is simply due to selection biases or redshift-dependent effects, we plot the ``actual'' $M_{2500\, \text{\AA}}$ values measured from our \textsc{PyQSOFit} continuum fluxes against our ``predicted'' $M_{2500\, \text{\AA}}$ values from Equation~\ref{eq:mumodel2} in narrow redshift bins in Figure~\ref{fig:predicted_Mi_binned}, using our best-fit flat~$w_0w_a$CDM model parameters to compute the ``actual'' and ``predicted'' absolute magnitudes. We find that our predicted 2500\,\AA\, absolute magnitudes track well with the actual absolute magnitudes, despite the range in apparent magnitudes in each bin. Therefore, we conclude that there is no strong redshift or luminosity dependence. This test demonstrates the validity of the standardization in our sample across redshifts.

\subsubsection{Internal reddening}

We expect intrinsic changes in the spectrum of the accretion disk with luminosity to be captured by our luminosity relation. However, a possible systematic could arise from variations in internal reddening with redshift and/or luminosity. AGNs with significant internal reddening from dust in the torus will have over-estimated intrinsic UV apparent magnitudes. To assess the impact of this internal reddening, we compute the X-ray to UV luminosity ratio parameter, $\alpha_{\rm{OX}} = -\log( L_{2~{\rm keV}}/L_{2500\, \text{\AA}})/2.605$ (e.g., \citep{Just2007}), where $L_{2~{\rm keV}}$ is the monochromatic $2$ keV X-ray luminosity. We use the $2{-}10$ keV luminosities from version 2.1 of the Chandra Source Catalog \citep{Evans2024} assuming our best-fitting $w_0w_a$CDM cosmology, and convert them to monochromatic 2~keV luminosities using $L_{2~{\rm keV}}
=
L_{2-10~{\rm keV}}\,
\frac{2-\Gamma}{10^{\,2-\Gamma}-2^{\,2-\Gamma}}
\,2^{\,1-\Gamma}$. We assume an X-ray spectral index of $\Gamma=1.9$, but do not apply a formal $K$-correction to the X-ray luminosities for consistency with \citep{Risaliti2015}.

\begin{figure}[!ht]
\centering
\includegraphics[width=0.8\textwidth]{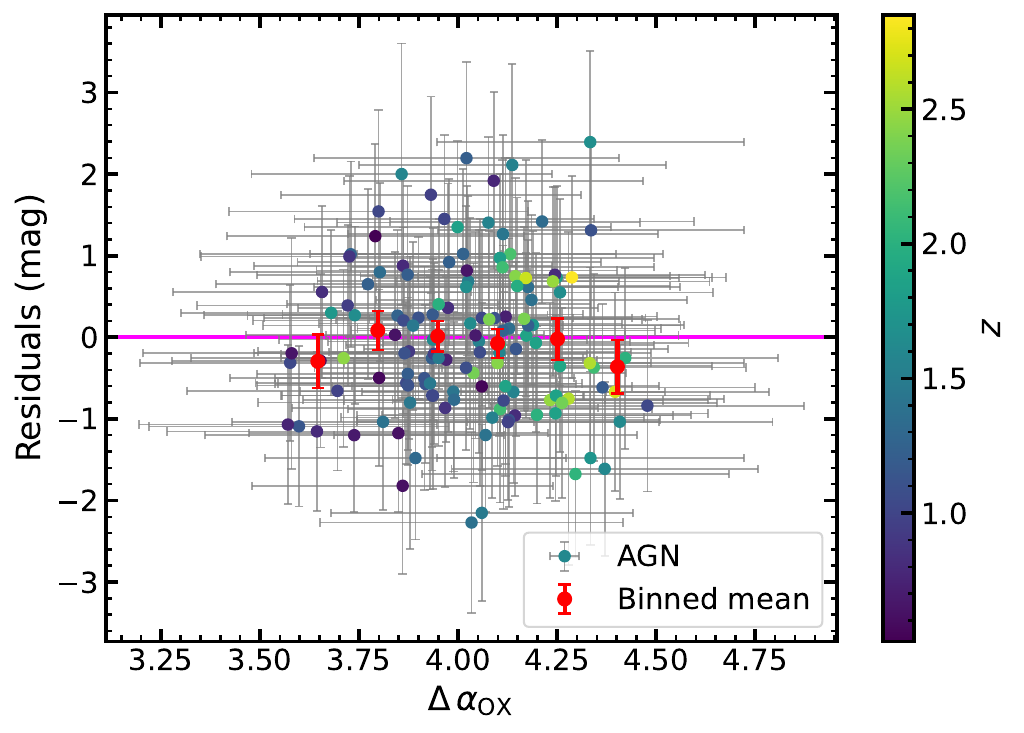}
\caption{AGN residuals on the Hubble diagram plotted against the actual minus the predicted X-ray–to–UV ratio, $\Delta\,\alpha_{\mathrm{OX}}$. Points are color-coded by redshift as indicated by the color bar. The solid magenta line marks zero residual. No significant trend is seen with $\Delta\,\alpha_{\mathrm{OX}}$, confirming that variations in the X-ray–to–UV ratio due to internal reddening do not introduce systematic biases in the standardization.\label{fig:alphaOx_residuals}}
\end{figure}

Type 1 AGN X-ray luminosities are relatively unaffected by dust reddening, so we would expect to see residual trends with $\Delta \alpha_{\rm{OX}}$ (difference between the actual and predicted $\alpha_{\rm{OX}}$ using the relation of \citep{Just2007}) if systematic errors from reddening are present. We plot the Hubble diagram residuals vs. $\alpha_{\rm{OX}}$ in Figure~\ref{fig:alphaOx_residuals}. We find no significant residual trends with $\alpha_{\rm{OX}}$, which indicates that variations in internal reddening, while likely to contribute to the intrinsic scatter in our relation, are unlikely to have a systematic impact on our cosmological inference. It is perhaps not surprising that our standardization is insensitive to reddening, because we compare the variability and apparent magnitudes at the same rest wavelength. This contrasts with previous work comparing X-ray to UV emission in AGNs (e.g., \citep{Lusso2025}).

\subsection{Physical origin of the luminosity--variability relation}

The anti-correlation between UV variability driving power and bolometric luminosity in active galactic nuclei (AGNs) may reflect luminosity-dependent changes in the physical conditions and geometry of the accretion disk. In lower-luminosity systems, UV emission arises at smaller radii where thermal timescales are short and magnetorotational instabilities \citep{Balbus1991} can drive strong stochastic fluctuations. As the bolometric luminosity increases due to higher accretion rates, the UV-emitting region shifts outward to larger radii \citep{Davis2011}, where dynamical and thermal timescales are longer.

The anti-correlation between UV variability driving power and bolometric luminosity may also partly arise from changes in the coherence of variability across the accretion disk. If the disk is composed of inhomogeneous emitting zones of fixed physical size, then more luminous AGNs with larger disks would exhibit more spatial averaging and thus more coherent, lower-amplitude fluctuations \citep{Dexter2011}. Alternatively, if variability originates at a localized region in the disk and propagates outward via Alfvén waves, then larger disks would exhibit diminished coherence and lower variability power at a given observing band due to increased propagation timescales and damping effects. The latter interpretation may be consistent with recent observational evidence of propagating waves of temperature perturbations that move across the disk \citep{Neustadt2022,Stone2023}. 

An additional contributor could be the varying role of X-ray reprocessing. In lower-luminosity AGNs, the X-ray corona can contribute a larger fraction of the total bolometric output (e.g., \citep{Lusso2016}), and fluctuations in the X-ray emission can efficiently irradiate and modulate the UV-emitting regions of the disk. This leads to enhanced short-timescale UV variability driven by coronal variability, consistent with disk reverberation signatures observed in lower-luminosity systems (e.g., \cite{Edelson2019,Cackett2020}). In contrast, more luminous AGNs tend to exhibit relatively weaker X-ray emission and reduced corona-disk coupling, thereby diminishing the impact of reprocessed X-ray variability on the UV light curves. This transition might reflect a change in the inner accretion structure or a relative suppression of the coronal heating mechanism at high accretion rates, possibly leading to deviations from standard thin disks (e.g., \citep{Wang2014}).

%\subsection{Error budget}
%The total uncertainty in the distance modulus ($\mu_{\mathrm{err}}$) for AGN is computed by combining observational, model-dependent, and astrophysical sources of uncertainty in quadrature. The contributing terms are as follows:

\pagebreak

\backmatter

\section{Supplementary information}

\begin{sidewaystable*}[p]

  \centering
  \begin{minipage}{\linewidth}
    \centering
    \caption{Summary of AGN properties and inferred distance moduli. The complete machine-readable table is available online.}
    \label{tab:agn_fits_table}

    \scriptsize
    \setlength{\tabcolsep}{2pt}
    \renewcommand{\arraystretch}{1.4}
    \begin{tabular}{@{}lccccccccc@{}}
\hline\hline
\textbf{SDSS Name} & RA & Dec & $z$ & $m_{2500}$ & $m_{2500}^{\mathrm{uncorr}}$ & $\mu$ & $\log\tau_{\mathrm{UV,RF}}$ & $\log\sigma_{\mathrm{UV}}$ & $\mathrm{Cov}(\log\sigma_{\mathrm{UV}},\,\log\tau_{\mathrm{UV,RF}})$ \\
& (deg) & (deg) &  & (mag) & (mag) & (mag) & (days) & (mag) &  \\\hline
\textbf{J011307.50$-$002309.4} & $18.2813$ & $-0.3860$ & $0.4549 \pm 0.0004$ & $21.44 \pm 0.02$ & $22.19 \pm 0.02$ & $43.81 \pm 1.22$ & $2.98 \pm 0.34$ & $-0.67 \pm 0.17$ & $0.028$ \\
\textbf{J235645.71$-$010801.8} & $359.1905$ & $-1.1339$ & $0.6996 \pm 0.0005$ & $21.88 \pm 0.02$ & $21.92 \pm 0.02$ & $42.90 \pm 1.41$ & $2.48 \pm 0.67$ & $-0.58 \pm 0.12$ & $0.017$ \\
\textbf{J212252.87+010900.8} & $320.7203$ & $+1.1502$ & $0.9437 \pm 0.0010$ & $22.43 \pm 0.03$ & $21.84 \pm 0.03$ & $43.54 \pm 0.99$ & $3.50 \pm 0.26$ & $-0.34 \pm 0.14$ & $0.029$ \\
\textbf{J015859.25$-$000114.1} & $29.7469$ & $-0.0206$ & $1.0222 \pm 0.0005$ & $23.11 \pm 0.03$ & $22.50 \pm 0.03$ & $43.80 \pm 0.96$ & $2.50 \pm 0.31$ & $-0.52 \pm 0.11$ & $0.023$ \\
\textbf{J230332.16$-$004421.4} & $345.8840$ & $-0.7393$ & $1.0233 \pm 0.0005$ & $22.53 \pm 0.02$ & $22.14 \pm 0.02$ & $45.45 \pm 1.16$ & $2.78 \pm 0.45$ & $-0.81 \pm 0.17$ & $0.046$ \\
\textbf{J000901.84+001146.6} & $2.2577$ & $+0.1963$ & $1.0434 \pm 0.0020$ & $21.86 \pm 0.02$ & $21.50 \pm 0.02$ & $43.30 \pm 0.95$ & $2.85 \pm 0.28$ & $-0.56 \pm 0.12$ & $0.028$ \\
\textbf{J225937.63$-$004204.3} & $344.9068$ & $-0.7012$ & $1.0545 \pm 0.0024$ & $22.51 \pm 0.02$ & $21.94 \pm 0.02$ & $42.93 \pm 0.98$ & $2.89 \pm 0.25$ & $-0.39 \pm 0.11$ & $0.019$ \\
\textbf{J003410.01+011350.5} & $8.5417$ & $+1.2307$ & $1.0908 \pm 0.0020$ & $21.20 \pm 0.02$ & $21.11 \pm 0.02$ & $42.83 \pm 1.05$ & $3.48 \pm 0.26$ & $-0.43 \pm 0.15$ & $0.029$ \\
\textbf{J004730.03+003120.3} & $11.8751$ & $+0.5223$ & $1.1398 \pm 0.0022$ & $23.00 \pm 0.03$ & $22.38 \pm 0.03$ & $45.08 \pm 1.22$ & $3.29 \pm 0.38$ & $-0.55 \pm 0.17$ & $0.036$ \\
\textbf{J212331.51$-$001742.3} & $320.8813$ & $-0.2951$ & $1.1565 \pm 0.0009$ & $22.37 \pm 0.02$ & $21.79 \pm 0.02$ & $44.23 \pm 1.01$ & $3.18 \pm 0.35$ & $-0.54 \pm 0.16$ & $0.048$ \\
\textbf{J230926.24$-$001044.8} & $347.3594$ & $-0.1791$ & $1.1695 \pm 0.0009$ & $22.36 \pm 0.02$ & $21.78 \pm 0.02$ & $44.11 \pm 0.94$ & $2.68 \pm 0.21$ & $-0.65 \pm 0.09$ & $0.012$ \\
\textbf{J013449.35+010302.2} & $23.7057$ & $+1.0506$ & $1.2348 \pm 0.0021$ & $22.10 \pm 0.02$ & $21.73 \pm 0.02$ & $44.75 \pm 1.01$ & $3.05 \pm 0.39$ & $-0.70 \pm 0.14$ & $0.042$ \\
\textbf{J211843.66$-$002122.0} & $319.6819$ & $-0.3561$ & $1.3650 \pm 0.0023$ & $22.87 \pm 0.02$ & $22.14 \pm 0.02$ & $44.28 \pm 0.94$ & $2.99 \pm 0.16$ & $-0.52 \pm 0.09$ & $0.008$ \\
\textbf{J232147.25+001853.2} & $350.4469$ & $+0.3148$ & $1.3738 \pm 0.0021$ & $21.54 \pm 0.02$ & $21.30 \pm 0.02$ & $42.53 \pm 1.53$ & $2.01 \pm 0.52$ & $-0.69 \pm 0.11$ & $-0.025$ \\
\textbf{J234433.06$-$001123.0} & $356.1378$ & $-0.1897$ & $1.4708 \pm 0.0021$ & $19.85 \pm 0.03$ & $20.55 \pm 0.03$ & $44.52 \pm 1.16$ & $3.40 \pm 0.38$ & $-0.93 \pm 0.12$ & $0.014$ \\
\textbf{J230527.69$-$003634.5} & $346.3654$ & $-0.6096$ & $1.4904 \pm 0.0027$ & $22.97 \pm 0.03$ & $22.13 \pm 0.03$ & $45.69 \pm 1.05$ & $2.91 \pm 0.28$ & $-0.75 \pm 0.13$ & $0.024$ \\
\textbf{J022536.29$-$002029.6} & $36.4012$ & $-0.3416$ & $1.5406 \pm 0.0022$ & $22.47 \pm 0.03$ & $21.79 \pm 0.03$ & $45.16 \pm 0.97$ & $2.77 \pm 0.29$ & $-0.77 \pm 0.12$ & $0.026$ \\
\textbf{J010625.24+002206.0} & $16.6052$ & $+0.3684$ & $1.5839 \pm 0.0038$ & $23.07 \pm 0.02$ & $22.20 \pm 0.02$ & $45.05 \pm 1.00$ & $2.32 \pm 0.30$ & $-0.77 \pm 0.09$ & $0.010$ \\
\textbf{J235922.96$-$004159.2} & $359.8457$ & $-0.6998$ & $1.6251 \pm 0.0036$ & $23.29 \pm 0.02$ & $22.34 \pm 0.02$ & $44.34 \pm 0.98$ & $2.80 \pm 0.22$ & $-0.51 \pm 0.11$ & $0.014$ \\
\textbf{J001548.25+000827.1} & $3.9511$ & $+0.1409$ & $1.7119 \pm 0.0046$ & $22.05 \pm 0.04$ & $21.59 \pm 0.04$ & $44.61 \pm 1.07$ & $2.90 \pm 0.41$ & $-0.72 \pm 0.18$ & $0.065$ \\
\textbf{J022514.57+000544.9} & $36.3107$ & $+0.0958$ & $1.8321 \pm 0.0070$ & $23.02 \pm 0.02$ & $22.11 \pm 0.02$ & $44.82 \pm 1.01$ & $2.59 \pm 0.32$ & $-0.68 \pm 0.13$ & $0.027$ \\
\textbf{J024037.61$-$005637.9} & $40.1567$ & $-0.9439$ & $1.8340 \pm 0.0029$ & $21.55 \pm 0.02$ & $21.21 \pm 0.02$ & $44.21 \pm 1.15$ & $3.26 \pm 0.32$ & $-0.65 \pm 0.17$ & $0.040$ \\
\textbf{J213311.55+010356.7} & $323.2982$ & $+1.0658$ & $1.8784 \pm 0.0040$ & $22.69 \pm 0.03$ & $21.93 \pm 0.03$ & $44.58 \pm 1.08$ & $2.87 \pm 0.38$ & $-0.62 \pm 0.16$ & $0.048$ \\
\textbf{J022655.43+005555.5} & $36.7310$ & $+0.9321$ & $2.1846 \pm 0.0029$ & $23.04 \pm 0.04$ & $21.99 \pm 0.04$ & $45.56 \pm 1.27$ & $2.94 \pm 0.52$ & $-0.71 \pm 0.19$ & $0.056$ \\
\textbf{J221018.78+005652.0} & $332.5783$ & $+0.9478$ & $2.1937 \pm 0.0026$ & $23.87 \pm 0.02$ & $22.37 \pm 0.02$ & $45.64 \pm 1.28$ & $2.46 \pm 0.53$ & $-0.70 \pm 0.13$ & $0.020$ \\
\textbf{J010934.56+004820.4} & $17.3940$ & $+0.8057$ & $2.4250 \pm 0.0030$ & $22.18 \pm 0.03$ & $21.51 \pm 0.03$ & $46.10 \pm 1.17$ & $2.79 \pm 0.45$ & $-0.96 \pm 0.19$ & $0.061$ \\
\textbf{J005503.61+000902.0} & $13.7650$ & $+0.1506$ & $2.4816 \pm 0.0047$ & $23.33 \pm 0.03$ & $22.18 \pm 0.03$ & $46.11 \pm 1.17$ & $2.77 \pm 0.38$ & $-0.79 \pm 0.17$ & $0.044$ \\
\textbf{J025130.43+011231.3} & $42.8768$ & $+1.2087$ & $2.6994 \pm 0.0052$ & $23.64 \pm 0.05$ & $22.19 \pm 0.05$ & $46.74 \pm 1.19$ & $2.81 \pm 0.40$ & $-0.83 \pm 0.18$ & $0.049$ \\
\textbf{J235142.11+000653.6} & $357.9255$ & $+0.1149$ & $3.0525 \pm 0.0067$ & $24.02 \pm 0.01$ & $22.78 \pm 0.01$ & $46.67 \pm 1.77$ & $2.07 \pm 0.61$ & $-0.94 \pm 0.19$ & $-0.006$ \\
\textbf{J231444.09$-$000309.7} & $348.6837$ & $-0.0527$ & $3.4509 \pm 0.0172$ & $23.12 \pm 0.04$ & $21.59 \pm 0.04$ & $46.19 \pm 1.44$ & $2.75 \pm 0.42$ & $-0.84 \pm 0.22$ & $0.054$ \\
\hline
\end{tabular}%

\par\noindent\raggedright\footnotesize{Note}—\textbf{SDSS Name}: SDSS identifier; \textbf{RA}: Right ascension in degrees; \textbf{Dec}: declination in degrees; $\boldsymbol{z}$: Spectroscopic redshift (systemic redshift from \citep{Wu2022}); $\boldsymbol{m_{2500}}$: Rest-frame UV apparent AB magnitude, corrected for selection bias using the completeness function; $\boldsymbol{m_{2500}^{\mathrm{uncorr}}}$: uncorrected rest-frame UV apparent magnitude; $\boldsymbol{\mu}$: distance modulus inferred from joint likelihood, corrected for selection bias (for plotting purposes only -- do not use for cosmological inference); $\boldsymbol{\log\tau_{\mathrm{UV,RF}}}$: logarithm of the rest-frame UV damping timescale; $\boldsymbol{\log\sigma_{\mathrm{UV}}}$: logarithm of the UV variability amplitude; and $\boldsymbol{\mathrm{Cov}(\log\sigma_{\mathrm{UV}},\log\tau_{\mathrm{UV,RF}})}$: covariance between the logarithm of the UV variability amplitude and UV damping timescale. All uncertainties are $1\sigma$ and logarithms are 10 based.

  \end{minipage}
%\end{adjustbox}
\end{sidewaystable*}

\bmhead{Funding} C.J.B. is supported by an NSF Astronomy and Astrophysics Postdoctoral Fellowship under the award AST-2303803. This material is based on work supported by the National Science Foundation under Award No. 2303803. This research award to NSF is partially funded by a generous gift of Charles Simonyi to the NSF Division of Astronomical Sciences. The award is made in recognition of significant contributions to Rubin Observatory’s Legacy Survey of Space and Time. P.N. acknowledges support from the Gordon and Betty Moore Foundation (Grant \#8273.01) and the John Templeton Foundation (Grant \#62286) that fund the Black Hole Initiative (BHI) at Harvard University where she serves as one of the PIs. P.N. also acknowledges support from the John Templeton Foundation via Grant \#126613 and support from the Department of Energy via the grant DE-SC0017660. W.Y. acknowledges support from the Dunlap Institute for Astronomy \& Astrophysics at the University of Toronto. The material presented in this paper is based upon work supported by NASA under award No. 80NSSC25K0311 to I.D. under the NASA FINESST program.

\bmhead{Acknowledgements} We thank Tonima Ananna, Neven Caplar, Bożena Czerny, Xin Liu, Swayamtrupta Panda, Alessandro Peca, Claudio Ricci, Yue Shen, and Qiaoya Wu, Fan Zou, and the LSST AGN Science Collaboration for helpful comments/discussions. We thank the Yale Center for Research Computing for guidance and use of the research computing infrastructure.

Funding for SDSS and SDSS-II has been provided by the Alfred P. Sloan Foundation, the Participating Institutions, the National Science Foundation, the U.S. Department of Energy, the National Aeronautics and Space Administration, the Japanese Monbukagakusho, the Max Planck Society, and the Higher Education Funding Council for England. The SDSS Web Site is http://www.sdss.org/.

The SDSS is managed by the Astrophysical Research Consortium for the Participating Institutions. 
The participating Institutions are the American Museum of Natural History, Astrophysical Institute Potsdam, University of Basel, University of Cambridge, Case Western Reserve University, University of Chicago, Drexel University, Fermilab, the Institute for Advanced Study, the Japan Participation Group, Johns Hopkins University, the Joint Institute for Nuclear Astrophysics, the Kavli Institute for Particle Astrophysics and Cosmology, the Korean Scientist Group, the Chinese Academy of Sciences (LAMOST), Los Alamos National Laboratory, the Max-Planck-Institute for Astronomy (MPIA), the Max-Planck-Institute for Astrophysics (MPA), New Mexico State University, Ohio State University, University of Pittsburgh, University of Portsmouth, Princeton University, the United States Naval Observatory, and the University of Washington.

Funding for the Sloan Digital Sky Survey IV has been provided by the Alfred P. Sloan Foundation, the U.S. Department of Energy Office of Science, and the Participating Institutions. SDSS-IV acknowledges support and resources from the Center for High Performance Computing  at the University of Utah. The SDSS website is www.sdss.org.

SDSS-IV is managed by the Astrophysical Research Consortium for the Participating Institutions of the SDSS Collaboration including the Brazilian Participation Group, the Carnegie Institution for Science, Carnegie Mellon University, Center for Astrophysics | Harvard \& Smithsonian, the Chilean Participation Group, the French Participation Group, Instituto de Astrof\'isica de Canarias, The Johns Hopkins University, Kavli Institute for the Physics and Mathematics of the Universe (IPMU) / University of Tokyo, the Korean Participation Group, Lawrence Berkeley National Laboratory, Leibniz Institut f\"ur Astrophysik Potsdam (AIP),  Max-Planck-Institut f\"ur Astronomie (MPIA Heidelberg), Max-Planck-Institut f\"ur Astrophysik (MPA Garching), Max-Planck-Institut f\"ur Extraterrestrische Physik (MPE), National Astronomical Observatories of China, New Mexico State University, New York University, University of Notre Dame, Observat\'ario Nacional / MCTI, The Ohio State University, Pennsylvania State University, Shanghai Astronomical Observatory, United Kingdom Participation Group, Universidad Nacional Aut\'onoma de M\'exico, University of Arizona, University of Colorado Boulder, University of Oxford, University of Portsmouth, University of Utah, University of Virginia, University of Washington, University of Wisconsin, Vanderbilt University, and Yale University.

The Pan-STARRS1 Surveys (PS1) and the PS1 public science archive have been made possible through contributions by the Institute for Astronomy, the University of Hawaii, the Pan-STARRS Project Office, the Max-Planck Society and its participating institutes, the Max Planck Institute for Astronomy, Heidelberg and the Max Planck Institute for Extraterrestrial Physics, Garching, The Johns Hopkins University, Durham University, the University of Edinburgh, the Queen's University Belfast, the Harvard-Smithsonian Center for Astrophysics, the Las Cumbres Observatory Global Telescope Network Incorporated, the National Central University of Taiwan, the Space Telescope Science Institute, the National Aeronautics and Space Administration under Grant No. NNX08AR22G issued through the Planetary Science Division of the NASA Science Mission Directorate, the National Science Foundation Grant No. AST-1238877, the University of Maryland, Eotvos Lorand University (ELTE), the Los Alamos National Laboratory, and the Gordon and Betty Moore Foundation.

Based on observations obtained with the Samuel Oschin Telescope 48-inch and the 60-inch Telescope at the Palomar Observatory as part of the Zwicky Transient Facility project. ZTF is supported by the National Science Foundation under Grants No. AST-1440341 and AST-2034437 and a collaboration including current partners Caltech, IPAC, the Oskar Klein Center at Stockholm University, the University of Maryland, University of California, Berkeley, the University of Wisconsin at Milwaukee, University of Warwick, Ruhr University, Cornell University, Northwestern University and Drexel University. Operations are conducted by COO, IPAC, and UW.

\bmhead{Competing interests} The authors declare that they have no competing financial interests.

\bmhead{Data availability} The light curve data is publicly available through SDSS, PS1, and ZTF. The Panthen+ and SH0ES data are available at \url{https://github.com/PantheonPlusSH0ES/DataRelease}.

\bmhead{Code availability} The code to perform light curve fitting, cosmology inference, and generate the figures is available at {\color{blue} this github link (TBD)}. 

\bmhead{Author contribution} 
ID: Conceptualization \& Execution, Methodology, Writing—Original Draft, Visualization.
CJB: Conceptualization, Methodology and Development of the Analysis Framework, 
Data analysis, Writing—Original Draft, Visualization.
PN: Conceptualization, Methodology, Investigation, Funding Acquisition, Writing—Original Draft and Review, as well as Editing.
WY: Data Analysis, Software, Methodology, Contribution to Writing. 
All authors: Contributed to the writing and revision of the manuscript and approved the final version for submission. 

\begin{appendices}

\section{Completeness curves}
\label{sec:completeness}

\begin{figure}
\centering
\includegraphics[width=1\textwidth]{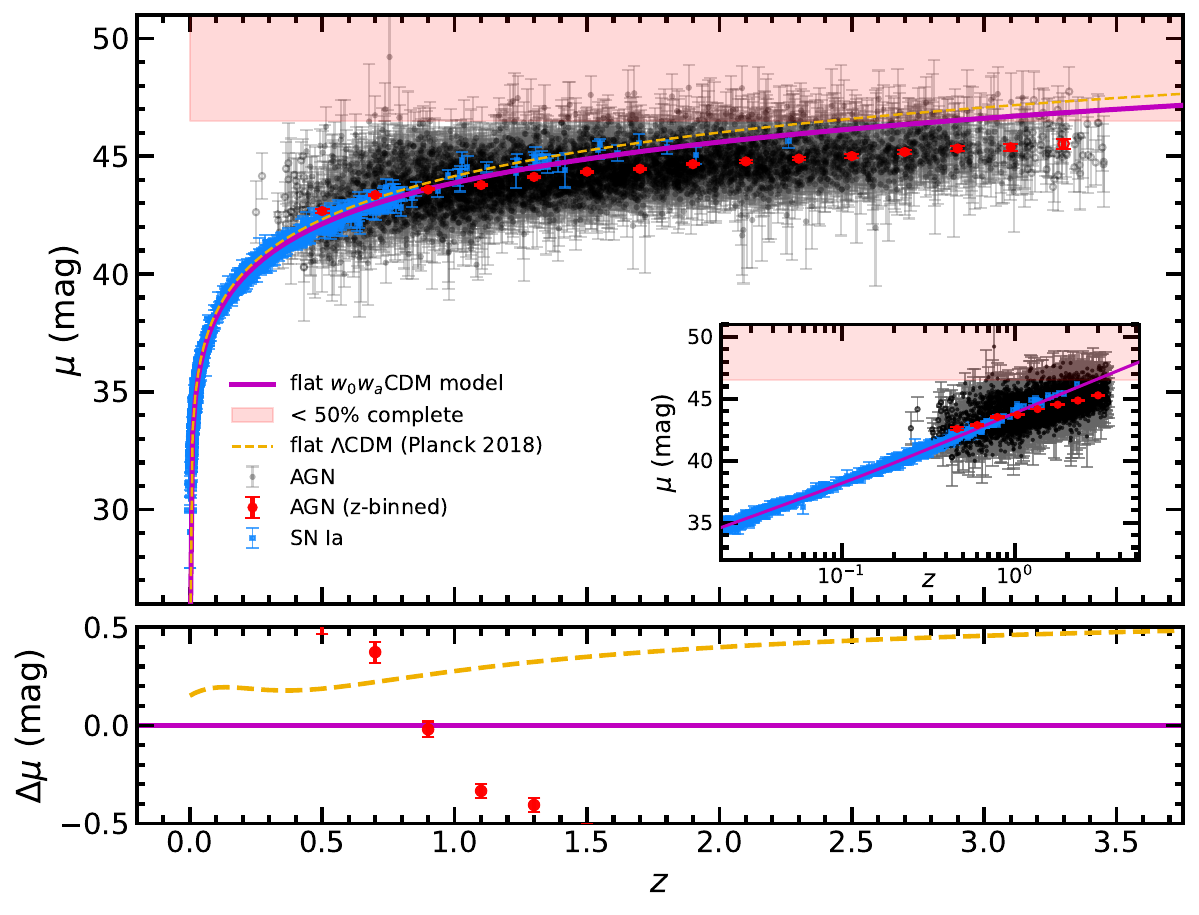}
\caption{Same as Figure~\ref{fig:hd} but with distance moduli uncorrected for incompleteness. The systematic offset between the fitted cosmological model and distance moduli are due to Malmquist bias.}
\end{figure}

\begin{figure}[!ht]
\centering
\includegraphics[width=0.8\textwidth]{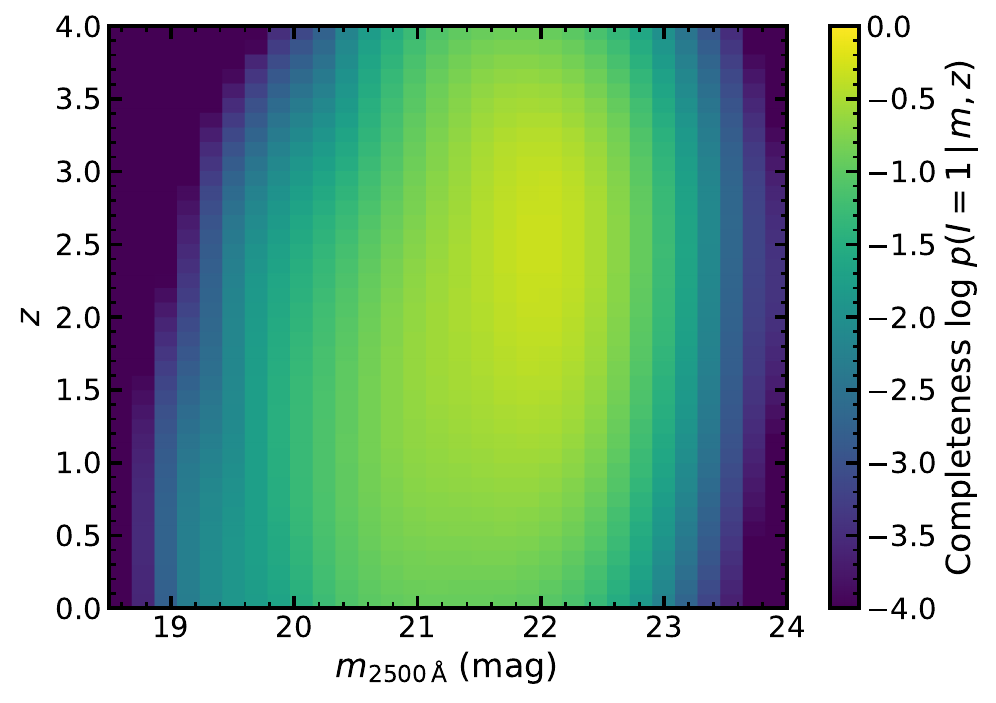}
\caption{Empirical completeness function $p(I{=}1|m, z)$ as a function of apparent magnitude in the $i$-band and redshift $z$. The completeness ratio is derived by comparing the observed apparent magnitude distribution to the true distribution. Yellow regions are more complete. This function is used to account for selection effects in the analysis. \label{fig:completeness_function}}
\end{figure}

To quantify the selection efficiency of our AGN sample as a function of apparent magnitude, we construct a completeness function $p(I{=}1|m,z)$. We begin by adopting the AGN luminosity function from \cite{Shen2020} to represent the intrinsic distribution of AGNs as a function of redshift and absolute magnitude. From this model, we generate a mock sample of AGNs over a redshift range of $0.5<z<3.5$ and the expected number of AGNs for a given apparent magnitude and comoving volume. We calculate $m_{i}(z{=}2)$ using the bolometric correction from \citep{Shen2011,Wu2022} and apply a $K$-correction following \cite{Richards2006} assuming a spectral index drawn from a normal distribution centered on $\alpha_\nu = -0.5$ and a width of $0.3$ \citep{Richards2006} to obtain $m_{2500\,\text{\AA}}$ (rest-frame), which is very close to the $i$ band central wavelength at $z=2$. We estimate the scatter in this conversion by calculating the standard deviation of the residuals, and convolve the selection function with a normal distribution with this scatter. This procedure yields an estimate of the true distribution of AGNs as a function of apparent magnitude and redshift in the absence of selection effects.

Next, we compare the modeled apparent magnitude distribution to the observed distribution of AGNs in our sample. We compute the ratio of the observed distribution to the modeled distribution in bins of magnitude and redshift. This ratio is the empirical completeness as a function of apparent magnitude and redshift. The resulting binned completeness curve is then fit using linear interpolation to derive the final completeness function that captures the varying selection efficiency resulting from the redshift-dependent SDSS spectroscopic AGN selection and our light curve quality cuts. The result is shown in Figure~\ref{fig:completeness_function}. This function is used as a weighting factor in Equation~\ref{eq:hubble_likelihood}, ensuring that our inferred cosmological parameters reflect the intrinsic distributions rather than the flux-limited sample. To assess the systematic errors from this procedure, we also used the X-ray luminosity function from \cite{Ananna2019} with the X-ray bolometric correction from \cite{Shen2020}. We assume an unobscured population with $20 < \log( N_{\rm{H}} / {\rm{cm}}^{-2}) < 22$, where $N_{\rm{H}}$ is the hydrogen column density. We found this does not change the results appreciably, and our inferred cosmological parameters are essentially unchanged to well within 1$\sigma$.

\section{Rest–frame wavelength dependence}
\begin{figure}[!ht]
\centering
\includegraphics[width=0.6\textwidth]{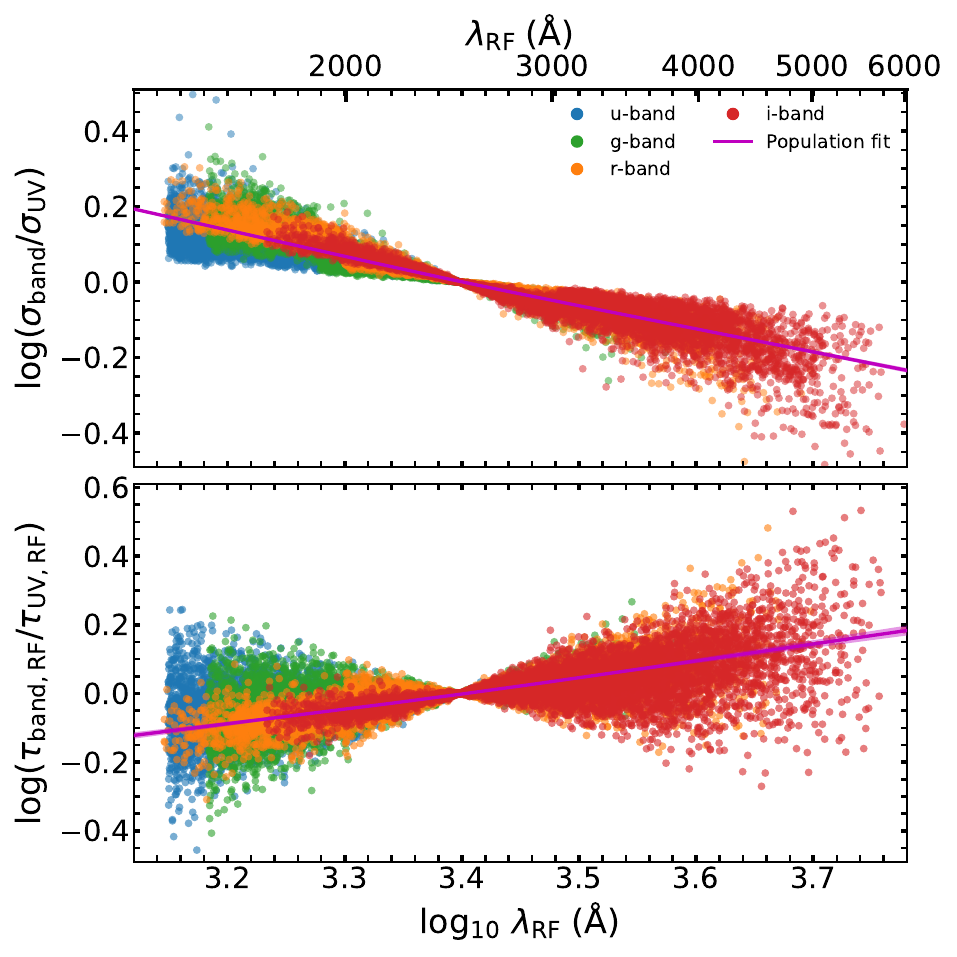}
\label{fig:wavelength_dependence}
\caption{\textbf{Rest–frame wavelength dependence of optical variability.} 
Trends of fractional variability amplitude $\sigma$ (top) and characteristic timescale $\tau$ (bottom) with rest–frame wavelength for the plotted AGN sample. 
Points are per–band measurements, color–coded by SDSS filter, expressed as ratios relative to the rest-frame UV quantities. 
The solid curve is the posterior–median population fit using the smoothly-broken power law (anchored at $2500$\,\AA); the shaded band is the $1\sigma$ credible region. Bands whose rest–frame blue edges fall shortward of Ly$\alpha$ ($1216$\,\AA) are excluded, producing the gap at the shortest wavelengths. }

\end{figure}

We investigate the rest–frame wavelength dependence of the variability amplitude and damping timescale by fitting a smoothly broken power law using \textsc{astropy}’s \texttt{SmoothlyBrokenPowerLaw1D} \citep{AstropyCollaboration2022} to these parameters in each band, with the break fixed at $2500\,\text{\AA}$. This model has the form $\sigma_{\rm{b}}/\sigma_{\rm{UV}} \propto \lambda_{\rm RF}^{\eta_{\sigma,\, \rm blue}}$ and $\tau_{\rm{b,\, RF}}/\tau_{\rm{UV,\, RF}} \propto \lambda_{\rm RF}^{\eta_{\tau,\, \rm blue}}$ at $\lambda_{\rm{RF}} < 2500$ \AA\ and $\sigma_{\rm{b}}/\sigma_{\rm{UV}} \propto \lambda_{\rm RF}^{\eta_{\sigma,\, \rm red}}$ and $\tau_{\rm{b,\, RF}}/\tau_{\rm{UV,\, RF}} \propto \lambda_{\rm RF}^{\eta_{\tau,\, \rm red}}$ at $\lambda_{\rm{RF}} > 2500$ \AA. We fit the per–band measurements with errors and propagate the quoted $1\sigma$ errors to the uncertainty bands in Fig.~\ref{fig:wavelength_dependence}. For the variability amplitude, the best–fit blue and red slopes are $\eta_{\sigma,\mathrm{blue}}=-0.70\pm0.02$ and $\eta_{\sigma,\mathrm{red}}=-0.61\pm0.02$, respectively. For the damping time, $\eta_{\tau,\mathrm{blue}}=0.42\pm0.04$ and $\eta_{\tau,\mathrm{red}}=0.49\pm0.04$, consistent with an unbroken power law over our wavelength range. These trends reflect the expected behavior from previous work, whereby variability weakens and characteristic timescales lengthen toward the red. Our results broadly agree with previous work. In particular, \citet{Yu2025} report $\eta_\sigma=-0.746\pm0.030$ and $\eta_\tau=+0.388\pm0.083$, while \citet{Stone2022} find $\eta_\sigma=-0.112^{+0.068}_{-0.070}$ and $\eta_\tau=+0.336^{+0.111}_{-0.109}$.

\section{Light curve length}
\label{appx:length}

\begin{figure}[!ht]
\centering
\includegraphics[width=\textwidth]{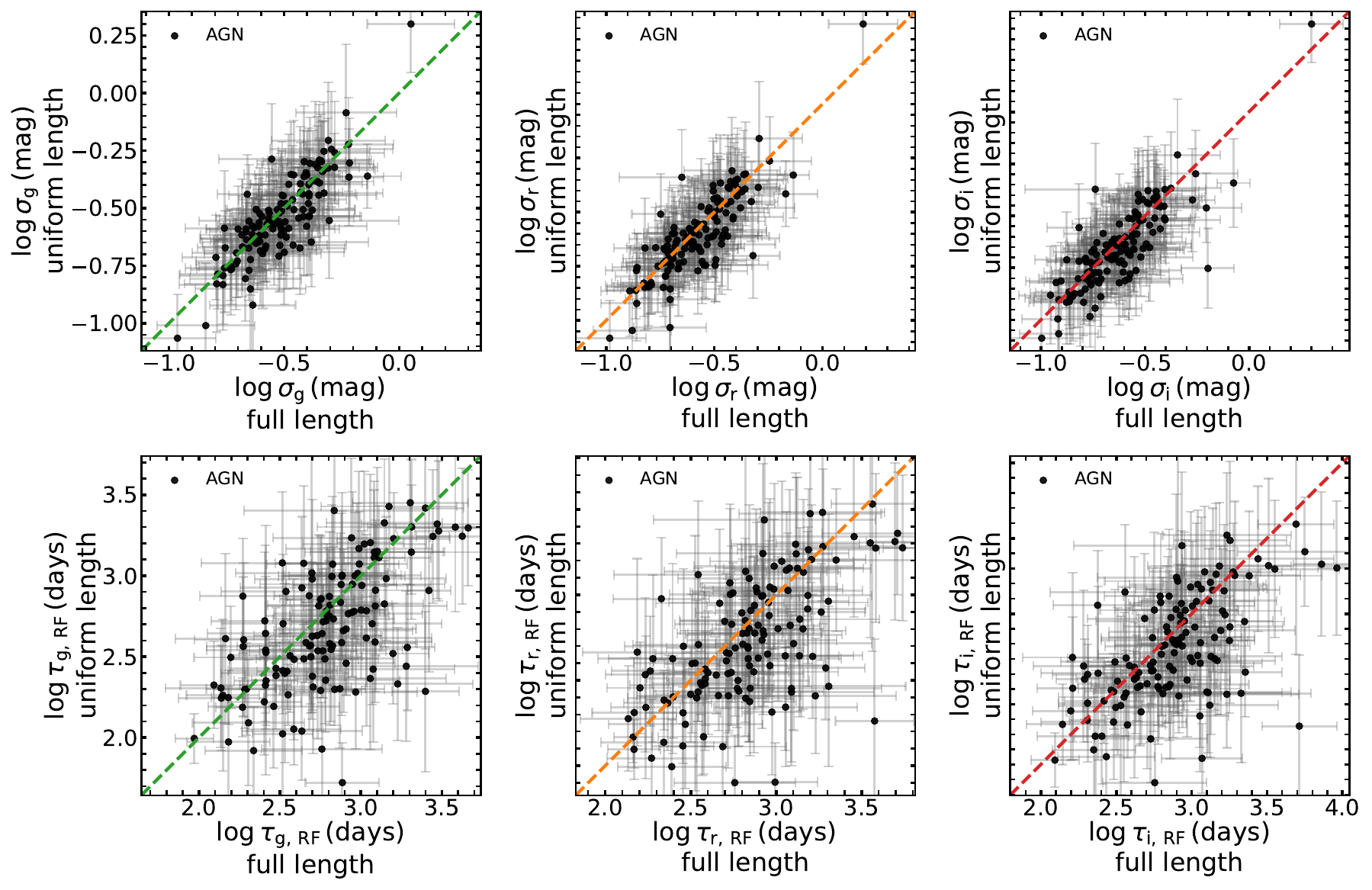}
\label{fig:length_test}
\caption{Comparison of variability amplitudes (top row) and damping timescales (bottom row) in $g$ (left column), $r$ (middle column), and $i$ (right columns) bands from this work with the full $\sim 20$ yr long light curves and the same light curves cut to a uniform length in the rest frame of $\sim 6.5$ yrs for a subset of our AGN matched with \citep{Stone2022}. This comparison includes the linear detrending mean function to subtract long timescale nonstationary features that was not performed by \citep{Stone2022}.} \label{fig:sigma_tau_same_length_poly1}
\end{figure}
\begin{figure}[!ht]
\centering
\includegraphics[width=\textwidth]{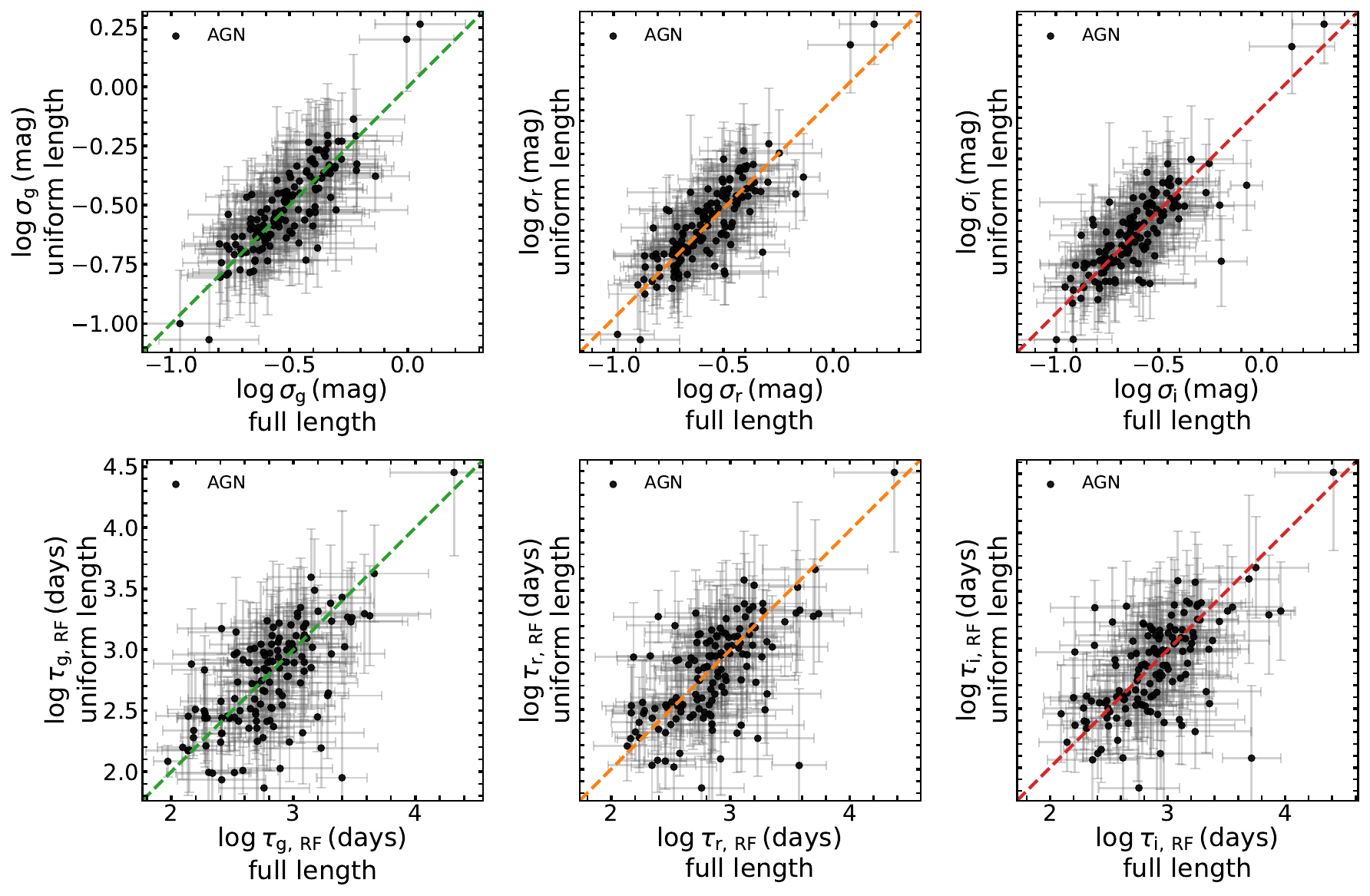}
%\label{fig:length_test}
\caption{Same as Figure~\ref{fig:sigma_tau_same_length_poly1} but without the linear detrending mean function.} \label{fig:sigma_tau_same_length_nopoly1}
\end{figure}

Maximum likelihood estimators are known to produce biased parameter estimates when the light curve length is shorter than $\sim 10$ times the damping timescale \citep{Kozlowski2017,Burke2021}. Using Bayesian sampling techniques, subsequent work demonstrated that the severity of the bias depends on the details of the prior distributions and parameter estimators adopted \citep{Hu2024}. In \textsc{EzTaoX}, the posterior median biases are $\sim 0.1$ dex when $\rho_{\rm{out}} = 1/5$ and 0.2 dex at $\rho_{\rm{out}} = 1$ using uniform priors in $\log \tau$ and $\log \sigma$ \citep{Yu2025eztaox}. In addition to these biases in the parameter estimates, \cite{Stone2022} found that the damping timescales became systematically longer using longer light curves of the same AGNs, even when the lengths of the light curves considered were all greater than 10 times the DRW timescale. \cite{Stone2022} attribute this to an unaccounted for nonstationary process that acts on timescales much longer than the damping timescale; a completely separate issue from the \citep{Kozlowski2017} bias. 

The lengths of the light curves in our sample vary from $\sim 2,000-8,000$ days in the rest frame depending on the redshift. Systematic differences in the recovered parameters due to varying rest-frame light curve lengths over redshift could influence our results across the Hubble diagram. To assess the potential bias in the inferred light curve fitting parameters introduced by varying rest-frame light curve lengths, we repeated our light curve fitting using a subset of AGNs with approximately uniform light curve durations. In this test, we clip the lengths of the light curves to $\sim 2,400$ days in the rest-frame at all redshifts. This choice of $\sim 2,400$ days results in much more consistent light curve lengths without dropping high-redshift AGNs from our sample. We perform this test with and without the linear detrending mean function (Equation~\ref{eqn:mean_detrending}). The results are shown in Figures \ref{fig:sigma_tau_same_length_poly1} and \ref{fig:sigma_tau_same_length_nopoly1}. We find that the inferred parameters are highly consistent between the uniform and varying rest-frame length datasets when the detrending mean function is used. Disabling the detrending term preserves overall consistency within uncertainties, with a slight weakening of agreement.

\section{Comparison with previous work}

\begin{figure}[!ht]
% nopoly1
\centering
\includegraphics[width=0.95\textwidth]{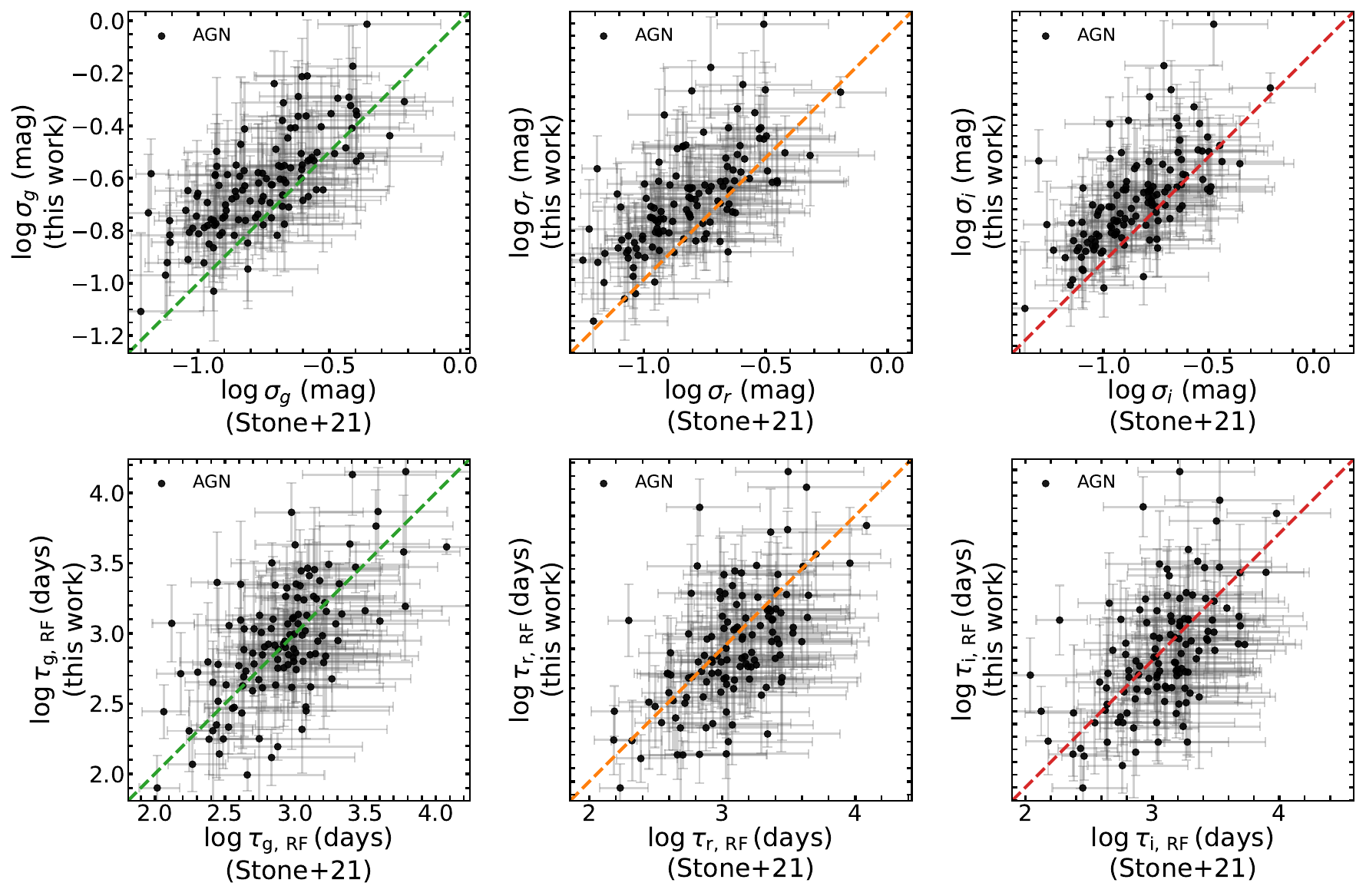}

\caption{Comparison of variability amplitudes (top row) and damping timescales (bottom row) in $g$ (left column), $r$ (middle column), and $i$ (right columns) bands from this work and the results of \cite{Stone2022}. This comparison does not include the linear detrending mean function to subtract long timescale nonstationary features, for better consistency with \citep{Stone2022} .\label{fig:stone_vs_this_work_nopoly1}}
\end{figure}

\begin{figure}[!ht]
\centering
\includegraphics[width=0.95\textwidth]{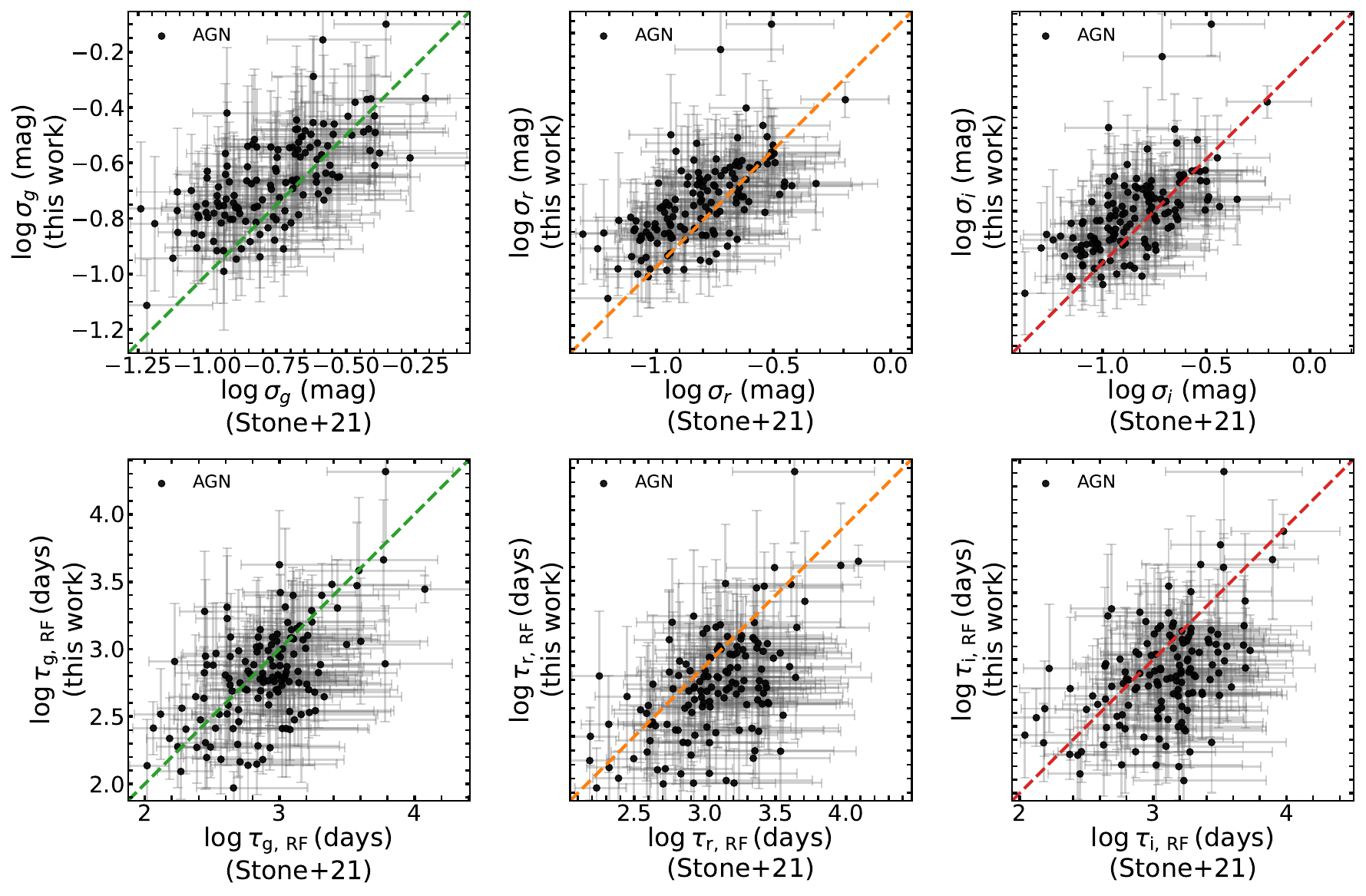}

\caption{Same as Figure~\ref{fig:stone_vs_this_work_nopoly1} but with the linear detrending mean function. The damping timescales from \citet{Stone2022} are longer than ours due to linear detrending. \label{fig:stone_vs_this_work_poly1}}
\end{figure}

As an additional consistency test, we compare our inferred parameters using our light curves with the DRW parameters inferred by \cite{Stone2022}. \cite{Stone2022} studied the variability of 190 SDSS AGN in Stripe 82 using $\sim 20$ year long $gri$ light curves from SDSS, PanSTARRS, and the Dark Energy Survey. These authors fitted DRWs to each band separately. We again perform this test with and without the linear detrending mean function. After our variability selection cuts, we find 188 common matches between our work and \cite{Stone2022}. The comparison is shown in Figures~\ref{fig:stone_vs_this_work_nopoly1} and
\ref{fig:stone_vs_this_work_poly1}. We find that our variability amplitudes are generally underestimated compared to \cite{Stone2022}, especially at low variability amplitudes, when no de-trending is done. With de-trending, our variability amplitudes are much better matched to the results of \cite{Stone2022}. Our damping timescales are generally consistent with the results of \cite{Stone2022} as well, but the $r$ and especially $i$ band timescales tend to be less correlated. We attribute this difference to our dataset having fewer points in $i$. 

\vspace{0.2in}

% poly1

\section{Broad line region contribution}

\begin{figure}[!ht]
\centering
\includegraphics[width=1\textwidth]{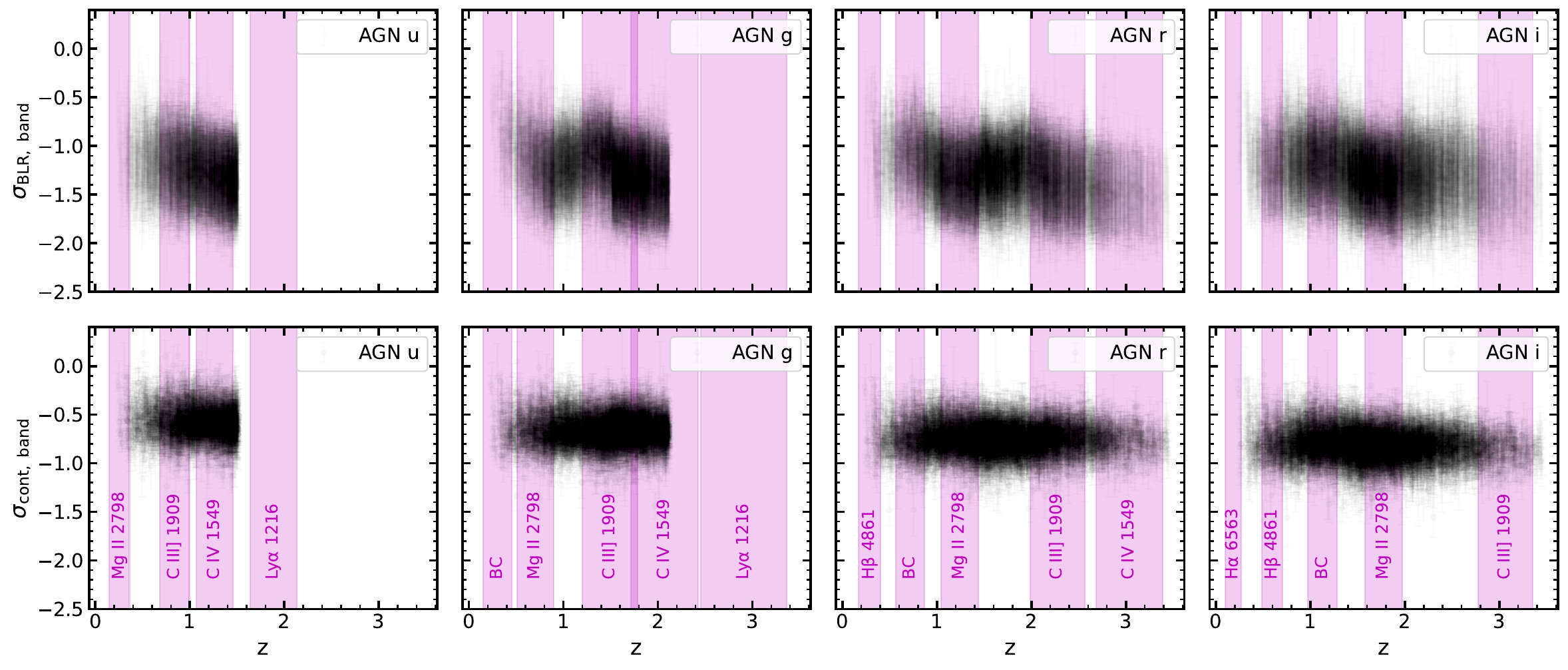}
\caption{\textbf{Variability amplitude as a function of redshift} in each band for the BLR (top panels) and continuum (bottom panels) for our AGN sample (black symbols). The locations of the major broad emission lines and the Balmer continuum (BC) are indicated on each panel. The apparent gaps and abrupt redshift edges arise from the exclusion of bands whose rest–frame blue edges fall shortward of Ly$\alpha$ (1216~\AA), where strong absorption and forest contamination suppress reliable continuum variability measurements.\label{fig:blr}}
\end{figure}

To check that our inference for the relative strength of the continuum vs. BLR contribution is reasonable, we plot the $\sigma_{\rm{BLR}}$ and  $\sigma_{\rm{cont}}$ in each band versus redshift in Figure~\ref{fig:blr}. We find that the BLR contribution ``wiggles'' with redshift as the different broad lines pass in and out of the observed frame filter bandpasses (e.g.,  \citep{Chelouche2014}). In particular, we see distinct bumps over the redshift range where the H$\alpha$, H$\beta$, and Balmer continuum emission is expected. In contrast, the continuum amplitude displays a subtle decreasing trend with redshift, as expected if the continuum variability amplitude anti-correlates with luminosity (since luminosity correlates with redshift). Our work reinforces the idea that photometric reverberation mapping may be possible with time domain surveys if reliable lags can be identified (e.g., \citep{Haas2011,PozoNunez2023,Czerny2023}).

%\section{Luminosity dependence}
%\begin{figure}
%\centering
%\includegraphics[width=1\textwidth]{figures/appendix/eta_params_vs_LOGLBOL_lbolbins_binned.png}
%\caption{}
%\end{figure}




\end{appendices}


\bibliography{bib}% common bib file
%\bibliographystyle{aasjournal}

\end{document}